\documentclass[a4paper,11pt]{article}
\pdfoutput=1
\usepackage{jcappub}
\usepackage{mathtools}
\usepackage{amsmath}
\usepackage{makecell}
\usepackage{booktabs}
\usepackage[table]{xcolor}

\newcommand{\gs}{g_\star}
\newcommand{\gss}{g_{\star s}}
\newcommand{\Trh}{T_\text{rh}}
\newcommand{\arh}{a_\text{rh}}
\newcommand{\Tmax}{T_\text{max}}
\newcommand{\rR}{\rho_R}
\newcommand{\rp}{\rho_\phi}
\newcommand{\Gp}{\Gamma_\phi}
\newcommand{\gammaE}{\gamma^2_\text{eff}}
\usepackage[normalem]{ulem}

\title{Two or three things particle physicists (mis)understand about (pre)heating}
\author[a]{Basabendu Barman,}
\author[b]{Nicolás Bernal,}
\author[c]{and Javier Rubio}
\affiliation[a]{\,\,Department of Physics, School of Engineering and Sciences, SRM University AP\\
Amaravati 522240, India}
\affiliation[b]{\,\,New York University Abu Dhabi\\
PO Box 129188, Saadiyat Island, Abu Dhabi, United Arab Emirates}
\affiliation[c]{\,\,Departamento de Física Teórica and\\
Instituto de Física de Partículas y del Cosmos (IPARCOS-UCM)\\
Universidad Complutense de Madrid, 28040 Madrid, Spain}
\emailAdd{basabendu.b@srmap.edu.in}
\emailAdd{nicolas.bernal@nyu.edu}
\emailAdd{javier.rubio@ucm.es}

\abstract{The transition from the end of inflation to a hot, thermal Universe, commonly referred to as (re)heating, is a critical yet often misunderstood phase in early Universe cosmology. This short review aims to provide a comprehensive, conceptually clear, and accessible introduction to the physics of (re)heating, tailored to the particle physics community. We critically examine the standard Boltzmann approach, emphasizing its limitations in capturing the intrinsically non-perturbative and non-linear dynamics that dominate the early stages of energy transfer. These include explosive particle production, inflaton fragmentation, turbulence, and thermalization; phenomena often overlooked in perturbative treatments. We survey a wide range of theoretical tools, from Boltzmann equations to lattice simulations, clarifying when each is applicable and highlighting scenarios where analytic control is still feasible. Special attention is given to model-dependent features such as (pre)heating, the role of fermions, gravitational couplings, and the impact of multifield dynamics. We also discuss exceptional cases, including Starobinsky-like models and instant (pre)heating, where (re)heating proceeds through analytically tractable channels without requiring full non-linear simulations. Ultimately, this review serves both as a practical guide and a cautionary tale, advocating for a more nuanced and physically accurate understanding of this pivotal epoch within the particle physics community.}
\begin{document}
\begin{flushright}
\end{flushright}
\maketitle
\section{Introduction}
Over the past decades, significant progress has been made in understanding the early Universe, with inflation emerging as a cornerstone of modern cosmology. Precision measurements of the cosmic microwave background and the successful predictions of Big Bang nucleosynthesis (BBN) offer compelling evidence for a thermal radiation-dominated phase in the early Universe. Yet, the transition from a cold, inflationary state to the hot Big Bang remains one of the least understood epochs in cosmology. This crucial phase—known as (re)heating\footnote{Throughout this review, we adopt the terms \textit{(re)heating} and \textit{(pre)heating}, rather than the more commonly used \textit{reheating} and \textit{preheating}, to emphasize that, within the standard inflationary scenario, the Universe need not have been in thermal equilibrium before the onset of inflation.}—marks the transfer of energy from the inflaton field to the Standard Model (SM) degrees of freedom, enabling the formation of matter and radiation as we know them.

Despite its pivotal role, (re)heating is often oversimplified in the particle physics literature. The commonly used Boltzmann equation (BEQ) framework, which models energy transfer via perturbative inflaton decays or scatterings, fails to account for key features of the process. These include explosive, non-perturbative effects such as parametric and tachyonic resonance, inflaton fragmentation, Bose enhancement, and turbulent thermalization—all of which can dramatically alter the dynamics. In particular, it has been shown that the Boltzmann and Bogoliubov approaches are inequivalent, especially at low-momentum modes where nonadiabatic particle production dominates (see, e.g., Ref.~\cite{Kaneta:2022gug}). In reality, the (re)heating phase is a rich multistage process. It begins with a non-perturbative particle production or (pre)heating stage, often requiring numerical lattice simulations to resolve the nonlinear evolution of coupled fields. As the system evolves, it gradually transitions to a more familiar perturbative regime where standard thermalization and decay processes can be described analytically. Both stages are sensitive to the specific inflationary potential, the type of interactions with matter, and the structure of the underlying theory.

The aim of this short review is to provide a pedagogical overview of (re)heating tailored to the particle physics community. Rather than attempting to cover the entire body of literature on the subject (a number of excellent reviews on this subject can already be found in the literature; see, e.g., Refs.~\cite{Bassett:2005xm, Allahverdi:2010xz, Amin:2014eta, Lozanov:2019jxc, Lozanov:2020zmy}), we focus on guiding the reader through the most essential concepts and mechanisms in a schematic and accessible manner. We demystify the limitations of standard perturbative approaches, highlight commonly overlooked non-linear phenomena, and clarify when simplified methods can still yield reliable insights. To this end, we survey a range of theoretical tools—including BEQs, semi-analytical estimates, and lattice simulations—emphasizing the specific conditions under which each framework is applicable. We also explore scenarios in which (re)heating proceeds solely via gravitational or scale-induced couplings, without requiring additional interactions with beyond-the-SM (BSM) fields.

The structure of the manuscript is as follows. In Section~\ref{sec:boltzmann}, we revisit the conventional Boltzmann-based approach to (re)heating and the assumptions behind it. Section~\ref{sec:limitation} explores the limitations of perturbation theory and introduces non-perturbative phenomena such as (pre)heating and inflaton fragmentation. We also highlight how the presence of fermions and multifield dynamics modifies (pre)heating and thermalization. Section~\ref{sec:nolattice} then focuses on the exceptional cases where full nonlinear simulations can be avoided and analytic control is still possible. We conclude in Section~\ref{sec:concl} by summarizing key takeaways and outlining open questions.

\section{The poor particle-physics approach} \label{sec:boltzmann}
The usual Boltzmann framework for studying (re)heating focuses on tracking the phase-space distribution of particles produced by perturbative decays or scatterings of the inflaton condensate by solving sets of BEQs~\cite{Kolb:1990vq}. This approach enables a systematic analysis of the time evolution of the energy densities associated with the inflaton, radiation, and any other exotic states generated during (re)heating. The Boltzmann framework has been extensively discussed in the literature, particularly within the particle physics community; see, e.g., Refs.~\cite{Albrecht:1982mp,Giudice:2003jh, Buchmuller:2004nz, Hahn-Woernle:2008tsk, Davidson:2008bu, Garcia:2017tuj, Bernal:2018qlk, Bernal:2018ins, Almeida:2018oid, Kaneta:2019zgw, Bernal:2019mhf, Garcia:2020wiy, Mambrini:2021zpp, Bernal:2021kaj, Barman:2021ugy, Kaneta:2021pyx, Clery:2021bwz, Haque:2022kez, Barman:2022tzk, Clery:2022wib, Bernal:2022wck, Haque:2023yra, Datta:2023pav, Silva-Malpartida:2023yks, Becker:2023tvd, Banerjee:2024caa, Barman:2024mqo, Barman:2024ujh, Bernal:2024ndy, Belanger:2024yoj}. Here, we provide a concise overview, emphasizing the key assumptions underlying this intrinsically perturbative treatment.

As is customary, we assume that the inflaton's self-interactions and its couplings to both the SM and the BSM fields are sufficiently small so as not to significantly alter the inflationary dynamics prior to the end of inflation. This cold-inflation assumption sets our analysis apart from warm-inflation scenarios~\cite{Berera:1995ie, Berera:1996nv, Berera:2008ar, Bastero-Gil:2009sdq, Bastero-Gil:2010dgy, Bastero-Gil:2016qru}, where dissipative effects play a central role throughout the inflationary period. Within this framework, any decay products generated during inflation are exponentially diluted by the rapid expansion of the Universe. Consequently, the inflaton field remains effectively homogeneous at the end of inflation, with residual inhomogeneities considered negligible. The post-inflationary evolution of the Universe is then primarily dictated by the inflaton potential, with minimal contributions from other energy components, until radiation domination takes over. Among the various inflationary models consistent with observations~\cite{Planck:2018jri}, ranging from the seminal Starobinsky model~\cite{Starobinsky:1980te, Starobinsky:1981vz, Starobinsky:1983zz, Kofman:1985aw} to Higgs inflation~\cite{Rubio:2018ogq} and its variants~\cite{Garcia-Bellido:2011kqb, Karananas:2016kyt, Casas:2017wjh, Casas:2018fum, Piani:2022gon} or other universal attractors~\cite{Kallosh:2013hoa, Artymowski:2016pjz}, we focus here for concreteness on a $T$-model of inflation with potential~\cite{Kallosh:2013hoa}
\begin{equation}\label{eq:inf-pot}
    V(\phi) = \lambda \, \Lambda^4\, \tanh^{n}\left(\frac{|\phi|}{\Lambda}\right) \simeq \lambda\, \Lambda^4 \times
    \begin{dcases}
        \left(\frac{\vert \phi\vert}{\Lambda}\right)^n &\text{ for } |\phi|\ll \Lambda\,,\\
        1 &\text{ for } |\phi|\gg \Lambda \,,
    \end{dcases}
\end{equation}
smoothly interpolating between a power-law behavior for small field values and a nearly constant plateau at large ones, with $\lambda$ a dimensionless coupling constant and $\Lambda$ an energy scale to be constrained from observations. With the usual chaotic slow-roll conditions~\cite{Linde:1983gd}, this asymptotic plateau allows for an accelerated expansion of the Universe that comes to an end when $\ddot a=0$, with $a$ the cosmological scale factor. The subsequent oscillatory evolution of the spatially homogeneous inflaton field around the minimum of its potential~\eqref{eq:inf-pot} is customarily described by an effective equation of motion
\begin{equation} \label{eq:eom0}
    \ddot\phi + (3\, H + \Gp)\, \dot\phi + V_{,\phi}(\phi) = 0\,,
\end{equation}
with the dots indicating derivatives with respect to coordinate time $t$, and the commas derivatives with respect to the field $\phi$. Here, $H$ represents the Hubble expansion rate, while $\Gp$ corresponds to the total inflaton decay rate, which quantifies the energy transfer rate from \textit{individual} inflaton quanta to the conventional matter and radiation components. This decay rate enters the equation of motion as a \textit{local} friction-like term, effectively governing the dissipation of the inflaton energy density over time. In most particle physics approaches, this decay rate is assumed to be either constant or only weakly dependent on time, allowing for a gradual and controlled energy transfer that smoothly connects inflationary (re)heating with the subsequent thermal history of the Universe. The exact form of the decay rate depends on the particular inflaton decay mode, which, in turn, is influenced by the underlying particle-physics model. 

In the context of (re)heating, the interactions of the inflaton with both SM and BSM fields can be broadly categorized into two primary mechanisms: direct decay and scatterings. On the one hand, in the decay channel, the inflaton field $\phi$ can transition, for instance, into pairs of lighter particles via trilinear interactions, such as $\mu\, \phi\, |\varphi|^2$ for scalar fields $\varphi$ (e.g., the Higgs boson doublet) or Yukawa-like couplings $y_\psi\, \phi\, \overline{\Psi}\, \Psi$ for Dirac fermionic fields $\Psi$. These interactions lead to two-body decays, with the coupling parameters $\mu$ and $y_\psi$ controlling their respective strengths. Depending on the underlying particle physics model, the main decay channel for the inflaton can alternatively involve other spin states or additional three or even four bodies, as discussed, for example, in Refs.~\cite{Ellis:2015kqa, Ellis:2015pla}. 
On the other hand, scattering processes, which provide an alternative pathway for energy transfer occur, for instance, via quartic couplings of the form $g\, \phi^2\, |\varphi|^2$, allowing 2-to-2 interactions of strength $g$ that populate the thermal bath with SM and BSM particles. Beyond this model-dependent channel, there exists an irreducible gravity-mediated production following from the unavoidable interaction term $\sqrt{-g}\, \mathcal{L}_{\rm grav} \supset -2/M_P\, h_{\mu \nu}\, T^{\mu \nu}$ between the graviton field $h_{\mu\nu}$ and the energy-momentum tensor $T_{\mu\nu}$ of all matter fields in the theory~\cite{Choi:1994ax}, with $M_P =(8\pi\, G_N)^{-1/2} \simeq 2.4\times 10^{18}$~GeV the reduced Planck mass. In this case, (re)heating can occur through 2-to-2 scatterings of the inflaton condensate into SM final states~\cite{Garny:2015sjg, Tang:2017hvq, Garny:2017kha, Bernal:2018qlk, Ahmed:2020fhc, Bernal:2021kaj, Haque:2022kez, Clery:2022wib, Co:2022bgh, Barman:2022qgt}, provided of course that the available energy density is large enough to partially overcome the involved Planck suppression. Alternatively, gravitational (re)heating can also occur through gravity-mediated decays as in the case of linear non-minimal interactions of the inflaton~\cite{Barman:2023opy}.

Given the above set of interactions and assuming that the effective masses of the decay products are smaller than that of the inflaton condensate,\footnote{Note in particular that this implicitly disregards potentially large time-dependent masses for the product particles induced by the inflaton field expectation value, as will be further emphasized in Section~\ref{sec:limitation}.} the inflaton's total decay width, computed using standard Feynman rules techniques, can be approximated as~\cite{Garcia:2020eof, Ahmed:2021fvt}
\begin{equation} \label{eq:inf-decay}
    \Gp =\frac{\gammaE}{8\pi} \times
    \begin{dcases}
        m_\phi(a) & \quad  \text{for decays into Dirac fermions},\\
        m^{-1}_\phi(a) & \quad \text{for decays into complex doublet scalars},\\
        \rho_\phi/m_\phi^3 & \quad \text{for annihilations into complex doublet scalars}, 
    \end{dcases}
\end{equation}
for explicit couplings to matter. In the case of minimal gravitational couplings~\cite{Clery:2021bwz, Clery:2022wib, Co:2022bgh}
\begin{equation} \label{Eq:ratephik}
    \Gamma_{\phi} \simeq \alpha_n\, M_P^5 \left(\frac{\rho_{\phi}}{M_P^4}\right)^{\frac{5n-2}{2n}},
\end{equation}
with $\alpha_n$ encoding the effect of oscillating inflaton condensate. Here, $\rho_\phi$ is the inflaton energy density, 
\begin{equation} \label{eq:inf-mass1}
    m_\phi^2(a) \equiv \frac{d^2V}{d\phi^2} \simeq n\, (n-1)\, \lambda^\frac{2}{n}\, \Lambda^\frac{2\, (4 - n)}{n} \rp(a)^{\frac{n-2}{n}}\simeq m_\phi(a_I) \left(\frac{a_I}{a}\right)^\frac{3 (n-2)}{n+2}
\end{equation}
stands for the inflaton mass in the vicinity of the origin ($\vert \phi\vert \ll \Lambda$), 
\begin{equation} \label{eq:geff}
    {\gammaE} = \gamma^2\, (n-1)\, (n+2) \left(\frac{\omega}{m_\phi}\right)^q\, \sum_{j=1}^\infty j^q\, |\mathcal{P}j|^2\, \left\langle \left[1 - \left(\frac{2\, m_p}{j\, \omega}\right)^2 \mathcal{P} \right]^{q/2} \right\rangle
\end{equation}
where $m_p$ is the mass of the product particle is the effective inflaton coupling to the daughter particles $\Psi$ and $\varphi$ after averaging over several inflaton oscillations~\cite{Ichikawa:2008ne, Kainulainen:2016vzv, Garcia:2020wiy, Clery:2021bwz, Clery:2022wib, Ahmed:2022tfm}, with $\gamma^2 = \{y_\psi^2,\, \mu^2,\, g^2n (n-1)\}$, $q= \{3,\, 1,\, 1\}$ and $P_j(t)$ the Fourier mode amplitudes of the oscillating inflaton condensate, defined as~\cite{Shtanov:1994ce, Ichikawa:2008ne} 
\begin{equation}
    \phi(t) = \phi_0(t)\, \mathcal{P}(t) = \phi_0(t) \sum_{j=1}^\infty \mathcal{P}_j(t)\, e^{-i\, j\, \omega\, t}\,, \hspace{10mm}     \omega = m_\phi\, \sqrt{\frac{n\, \pi}{2n-1}}\, \frac{\Gamma \left( \frac{n+1}{2n}\right)}{\Gamma \left( \frac{1}{2n}\right)}\,,
\end{equation}
where $\mathcal{P}(t)$ is a quasi-periodic, fast-oscillating function, and $\phi_0(t)$ is a slowly-varying envelope. Note that in Eq.~\eqref{eq:geff} the factor in the edgy brackets correspond to an average over the kinematical factors. For $n=2$, both the inflaton mass and the frequency $\omega$ in this mode decomposition are exactly constant, while displaying an explicit field dependence for $n>2$ which makes them decrease with time.

Under the assumption that the backreaction of the aforementioned relativistic decay products is small, the inflaton phase-space distribution function is given by some sharply peaked or delta function momentum distribution $f_\phi(k,\,t) = (2\pi)^3\, n_\phi(t)\, \delta^{(3)}(\vec{k})$, with $n_\phi$ the inflaton number density. The corresponding integrated BEQ for $n_\phi$ can be written as~\cite{Kolb:1990vq} 
\begin{equation}
    \dot n_\phi + 3\,H\,n_\phi = -\int d\Pi_\phi\, d\Pi_p \left[\left|\overline{\mathcal{M}}\right|^2_{\phi\to p} \,f_\phi \left(1 \pm f_p\right) - \left|\overline{\mathcal{M}}\right|^2_{p\to\phi}\, f_p \left(1 \pm f_\phi\right) \right],
\end{equation}
where $p$ collectively denotes all the final particles produced from inflaton decay or scattering, $d\Pi$'s are the Lorentz-invariant phase space, $|\overline{\mathcal{M}}|^2$ is the spin-averaged squared amplitude for the underlying process concerned (with any symmetry factors included), and $\left(1\pm f_p\right)$ takes care of Pauli-blocking/stimulated emission effects for the product particles. Clearly, all information from the underlying particle physics model is imprinted in the matrix element $\mathcal{M}$, which is computed perturbatively for a given decay/scattering diagram involving inflaton and SM fields. Now, assuming $i)$ no production of inflaton condensate from the bath, i.e., $|\overline{\mathcal{M}}|^2_{p\to\phi} = 0$, and $ii)$ that the Bose enhancement and Pauli blocking effects due to the decay products are negligible (that is, $1\pm f_p \simeq 1$), one can arrive at the evolution equation for the inflaton energy density $\rho_\phi$, which reads~\cite{Turner:1983he}
\begin{equation} \label{eq:drhodt}
    \frac{d\rp}{dt} + \frac{6\, n}{2 + n}\, H\, \rp = - \frac{2\, n}{2 + n}\, \Gp\, \rp\,.
\end{equation}
For small coupling constants compatible with the radiative instability of the inflaton potential during inflation, the term $H\, \rp$ associated with the cosmological expansion during (re)heating typically dominates over the interaction term $\Gp\, \rp$. In this regime, the evolution of the inflaton energy density and the associated Hubble rate $H\propto \rho_\phi^{1/2}$ evolve as 
\begin{equation} \label{eq:rpsol}
    \rp(a) \simeq \rp (\arh) \left(\frac{\arh}{a}\right)^\frac{6\, n}{n+2}, \quad\quad     H(a) \simeq H(\arh)  \left(\frac{\arh}{a}\right)^\frac{3\, n}{n + 2},
\end{equation}
with $a_I$ and $\arh$ the scale factors at the end of inflation and at the conclusion of the (re)heating stage, respectively. This scaling is associated to an effective equation-of-state~\cite{Turner:1983he, Johnson:2008se}
\begin{equation}\label{EoS}
    \langle w\rangle=\frac{\langle p_\phi\rangle}{\langle\rho_\phi\rangle}=\frac{n-2}{n+2}\,,    
\end{equation}
where $\langle...\rangle$ denotes an average over several inflaton oscillations. On the other hand, the evolution of the radiation energy density $\rR$ is governed by a BEQ of the form~\cite{Turner:1983he}
\begin{equation} \label{eq:dRdt}
    \frac{d\rR}{dt} + 4\, H\, \rR = + \frac{2\, n}{2 + n}\, \Gp\, \rp\,,
\end{equation}
which, using Eq.~\eqref{eq:rpsol}, admits a solution
\begin{equation} \label{eq:rR_int}
    \rR(a) \simeq \frac{2\, \sqrt{3}\, n}{2 + n}\, \frac{M_P}{a^4} \int_{a_I}^a \Gp(a')\, \sqrt{\rp(a')}\, a'^3\, da'\,.
\end{equation} 
Assuming that the decay products thermalize immediately after production,\footnote{For detailed analysis of (perturbative) thermalization see, e.g. Refs.~\cite{Davidson:2000er, Allahverdi:2002pu, Kurkela:2011ti, Harigaya:2013vwa}.} one can define an instantaneous radiation temperature as
\begin{equation} \label{rhoradiation}
    T(a)= \left(\frac{30 \,\rR(a)}{\pi^2\, \gs(a)} \right)^{1/4},
\end{equation}
with $\gs(a)$ the evolving number of relativistic degrees of freedom contributing to the SM energy density. Note that, in general, this temperature can significantly exceed the so-called \textit{(re)heating temperature} $\Trh \equiv T(\arh)$ marking the onset of radiation domination, $\rR(\arh) = \rp(\arh)$, facilitating with it key processes such as baryogenesis or the thermalization of weakly-interacting dark matter candidates with the radiation bath~\cite{Giudice:2000ex}. During (re)heating, the SM temperature typically follows as simple scaling
\begin{equation}
    T(a) \simeq \Trh \left(\frac{\arh}{a}\right)^\alpha,
\end{equation}
with $\alpha$ depending on the inflaton potential and the details fo the energy transfer (spin of the daughter particles and annihilation/decay channel). In any case, to ensure the success of BBN, the (re)heating temperature must satisfy $\Trh > T_\text{BBN} \simeq 4$~MeV~\cite{Sarkar:1995dd, Kawasaki:2000en, Hannestad:2004px, DeBernardis:2008zz, deSalas:2015glj, Hasegawa:2019jsa, Barbieri:2025moq}.

\begin{figure}[t!]
    \def\sepf{0.53}
    \centering
    \includegraphics[width=\sepf\columnwidth]{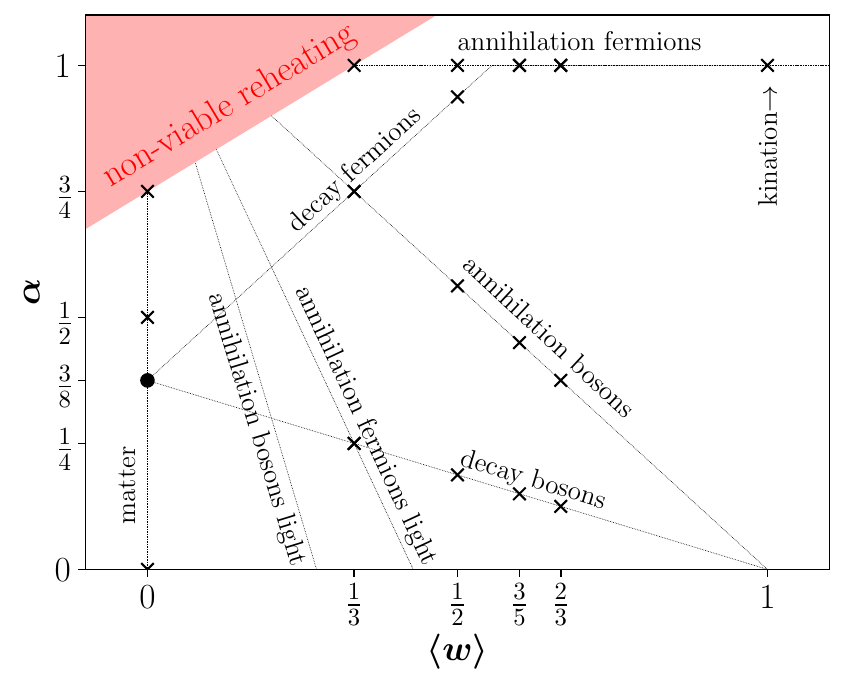}
    \caption{Summary of different (re)heating scenarios. The black dot corresponds to the case $n = 2$, where the inflaton scales as non-relativistic matter and decays into SM particles with a constant decay width, while the black crosses correspond to the alternative scenarios described in the text. The red area in the upper left corner does not give rise to viable (re)heating.}
    \label{fig:cosmo}
\end{figure} 
For the case $n = 2$, the inflaton scales as non-relativistic matter ($\langle w\rangle = 0$), leading to a temperature scaling of $\alpha = 3/8$, independently of the spin of the decay products~\cite{Giudice:2000ex}. However, for steeper potentials, the scaling depends on both the potential and the nature of the decay products: If the inflaton decays into bosons, one finds $\alpha = 3/(2(n+2))$, while for decays into fermions, $\alpha = \min\left[3(n-1)/(2(n+2)),\, 1\right]$. Alternatively, if the inflaton annihilates into bosons through contact interactions, the temperature scales as $\alpha = 9/(2(n+2))$ for $n \geq 3$. In scenarios where (re)heating occurs through $s$-channel annihilation mediated by a light scalar, resonant effects can modify the scaling to $\alpha = 3(7 - 2n)/(2(n+2))$ for bosonic final states, and $\alpha = 3(5 - n)/(2(n+2))$ for fermionic final states~\cite{Barman:2024mqo}. If the mediator is instead heavy, annihilation into fermions yields $\alpha = 1$~\cite{Barman:2024mqo}. Interestingly, (re)heating scenarios with constant temperature evolution, that is, $\alpha = 0$, are also possible~\cite{Co:2020xaf, Barman:2022tzk, Chowdhury:2023jft, Cosme:2024ndc}. Moreover, if the inflaton energy density redshifts faster than that of radiation (i.e., for $\langle w \rangle > 1/3$), it need not decay or annihilate at all. In such cases (such as kination) the radiation can eventually dominate with $\alpha = 1$~\cite{Spokoiny:1993kt, Ferreira:1997hj}. These various (re)heating scenarios are summarized in Fig.~\ref{fig:cosmo}, which maps the parameter space in the [$\langle w \rangle$, $\alpha$] plane. The black lines represent specific scenarios, illustrating the interplay between the inflaton dynamics and the thermal history. The vertical gray dotted line marks the case $\langle w \rangle = 0$, while the red-shaded region in the upper-left corner, where $\alpha \leq 3(1 + \langle w \rangle)/4$, corresponds to non-viable (re)heating, since in this regime the SM radiation energy density never overtakes that of the inflaton.

\begin{figure}[t!]
    \def\sepf{0.53}
    \centering    \includegraphics[width=\sepf\columnwidth]{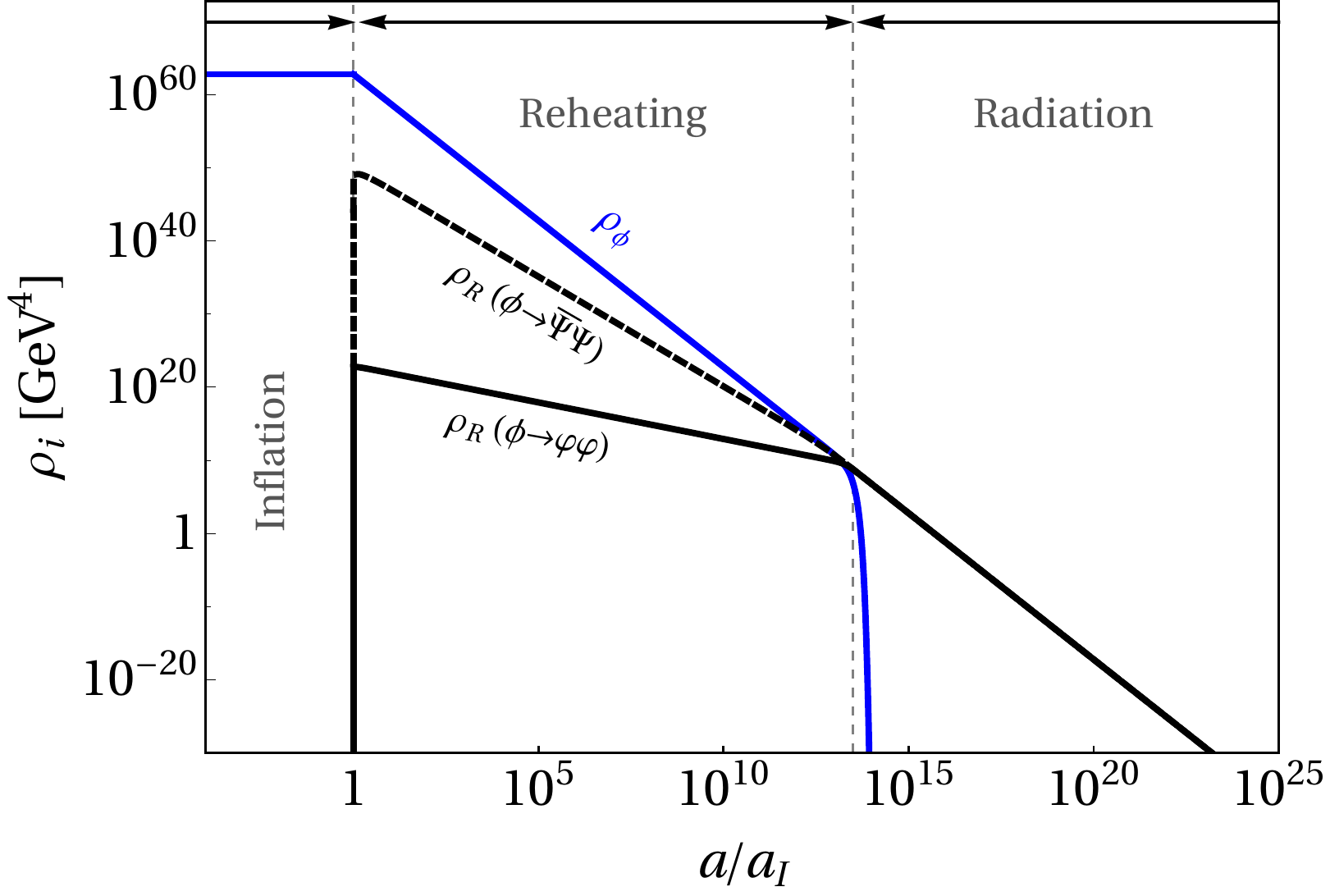}
    \caption{Evolution of the energy densities as a function of the scale factor $a$ for the inflaton ($\rp$, blue line) and the SM bath $\rR$, assuming an inflaton decays into fermions (dashed black) or scalars (solid black), for a quartic potential $n = 4$.}
    \label{fig:BEQ}
\end{figure} 
Figure~\ref{fig:BEQ} illustrates the evolution of energy densities as functions of the scale factor $a$ for the inflaton ($\rp$, blue line) and the SM radiation bath ($\rR$), assuming that the inflaton decays into fermions (dashed black) or scalars (solid black). The results correspond to a quartic inflaton potential, for which $n = 4$, and are obtained from a full numerical solution of the BEQs~\eqref{eq:drhodt} and~\eqref{eq:dRdt}. During (re)heating, the inflaton energy density scales as $\rp(a) \propto a^{-4}$, while the radiation energy density scales as $\rR(a) \propto a^{-3}$ in the case of fermionic decays, or as $\rR(a) \propto a^{-1}$ for scalar decays.

Finally, we note that beyond matter production, the (re)heating era is also a source of a potentially observable stochastic gravitational wave (GW) background, from channels that include: (i) graviton Bremsstrahlung~\cite{Nakayama:2018ptw, Huang:2019lgd, Barman:2023ymn, Barman:2023rpg, Kanemura:2023pnv, Bernal:2023wus, Tokareva:2023mrt, Hu:2024awd, Choi:2024acs, Barman:2024htg, Inui:2024wgj, Jiang:2024akb} and graviton pair production from inflaton annihilations~\cite{Ema:2015dka, Ema:2016hlw, Ema:2020ggo, Choi:2024ilx}, (ii) graviton emission from inflaton scattering with either thermalized~\cite{Xu:2024fjl} or non-thermalized~\cite{Bernal:2025lxp} decay products, and (iii) scattering among thermalized particles during reheating~\cite{Bernal:2024jim}. For GW production in the radiation-dominated era, see also~\cite{Ghiglieri:2015nfa, Ghiglieri:2020mhm, Ringwald:2020ist, Ringwald:2022xif, Klose:2022rxh, Ghiglieri:2022rfp, Drewes:2023oxg, Ghiglieri:2024ghm} for studies involving thermal plasma interactions.

\section{Here be dragons} \label{sec:limitation}
\begin{figure}[t!]
    \def\sepf{1}
    \centering
    \includegraphics[width=\sepf\columnwidth]{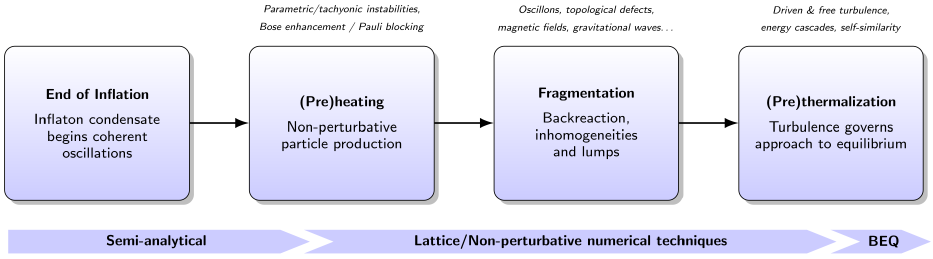}
    \caption{Sketch of the typical stages of post-inflationary (re)heating in oscillatory models of inflation, and the approximate computational regimes.}
    \label{fig:(re)heating-stages}
\end{figure} 
Almost three decades ago, Kofman, Linde, and Starobinsky~\cite{Kofman:1994rk, Kofman:1997yn} (see also Refs.~\cite{Traschen:1990sw, Dolgov:1989us, Shtanov:1994ce, Kaiser:1995fb}) identified fundamental shortcomings in the widely used Boltzmann approach presented in the previous section, demonstrating that it is not just inaccurate, but also fundamentally inadequate, at least during the initial stages of (re)heating; cf.~Fig.~\ref{fig:(re)heating-stages}. First, Eq.~\eqref{eq:eom0} violates the fluctuation-dissipation theorem, which states that dissipation within a system necessarily generates fluctuations~\cite{Kubo:1966fyg, Gleiser:1993ea}.  In particular, the analysis does not account for how these fluctuations modify the effective mass of the inflaton condensate. Furthermore, the use of a local decay term relies on the adiabatic-Markovian approximation, which is valid when the timescales associated with the decay process are much shorter than the Hubble time, which might not apply at  at the onset of reheating or more generically in warm inflationary scenarios (see e.g. Refs.~\cite{Berera:1998gx, Berera:2001gs}). Another issue is that the perturbative approximation employed when computing amplitudes and decay rates neglects both the \textit{coherent} nature of the inflaton field and crucial phenomena such as Bose enhancement and Pauli blocking effects. Most critically, even for coupling values where radiative corrections still remain negligible, perturbative methods break down. For example, if the inflaton oscillates in a potential steeper than quadratic, homogeneous oscillations can trigger explosive instabilities leading to rapid amplification of field fluctuations. Unlike the slow, steady energy leakage of perturbative decays, this mechanism violently disrupts the inflaton condensate, causing its fragmentation and initiating a highly non-linear interplay with its decay products, which can also undergo non-perturbative enhancements, further complicating the dynamics and often requiring the use of classical lattice simulations for a proper treatment; see e.g. Refs.~\cite{Felder:2000hq, Huang:2011gf, Figueroa:2021yhd}.\footnote{For a justification of the validity of the classical analysis and its limitations, see e.g. Refs.~\cite{Berges:2002cz, Berges:2013lsa, Tranberg:2023uzs}.} This section explores these dramatic, non-perturbative phenomena that challenge the conventional perturbative picture presented in Section~\ref{sec:boltzmann}.

\subsection{Inflaton self-resonance and fragmentation}
At the end of inflation, the inflaton field undergoes coherent oscillations around the minimum of its potential. In models where the potential is of the form of Eq.~\eqref{eq:inf-pot} with $n > 2$, these oscillations can trigger self-resonance instabilities, driving the energy transfer from the homogeneous condensate $\bar \phi(t)$ into spatial gradients and ultimately leading to its fragmentation.\footnote{Note that this implicitly disregards gravitational effects, which have been shown to ultimately lead to the formation of nonlinear structures, even in free field cases ($n = 2$), much like the gravitational instabilities seen in pressureless matter in the late Universe~\cite{Jedamzik:2010dq, Easther:2010mr}. \label{footnote}} This instability arises because the oscillating field sources its own perturbations $\delta\phi(t, \vec{x}) = \phi(t, \vec{x}) - \bar{\phi}(t)$ amplifying them non-perturbatively~\cite{Khlebnikov:1996mc, Greene:1997fu, Micha:2002ey, Micha:2004bv, Garcia:2023eol, Garcia:2023dyf}. In Fourier space,  we have 
\begin{equation}
    {\delta\ddot\phi}_{\vec{k}} + 3\, H\, \delta\dot \phi_{\vec{k}} + \left[\frac{k^2}{a^2} + V_{,\phi\phi}(\bar{\phi})\right] \delta\phi_{\vec{k}} = 0\,,
\end{equation}
with $V_{,\phi\phi}(\bar \phi)$ a periodic function of time.  The general solution to this equation, incorporating the expansion, takes the form~\cite{magnus2004hills}
\begin{equation}
    \delta\phi_k = \mathcal{P}_{k+}(t)\, e^{+\mu_k\, t} + \mathcal{P}_{k-}(t)\, e^{-\mu_k\, t},
\end{equation}
with $\mathcal{P}_{k\pm}(t)$ periodic functions of time determined by the initial conditions, and $\mu_k$ the so-called \textit{Floquet exponents}, which can change with time due to the presence of the expansion term. If the real part of these exponents does not vanish for specific $k$ values and $\operatorname{Re}(\mu_k) \gg H$, then an \textit{unstable solution} grows exponentially with time, indicating \textit{non-adiabatic} particle production. If efficient enough, this growth of perturbations eventually results in a backreaction on the condensate. As a result, the initially homogeneous field breaks into localized, transient structures, accelerating the energy transfer away from the homogeneous mode. At this stage, particle occupation numbers become poorly defined, and a nonlinear wave description is more suitable~\cite{Khlebnikov:1996mc, Greene:1997fu, Micha:2002ey, Micha:2004bv}. 

\begin{figure}
    \centering
    \begin{minipage}{0.3\textwidth}
        \centering
        \includegraphics[width=\linewidth]{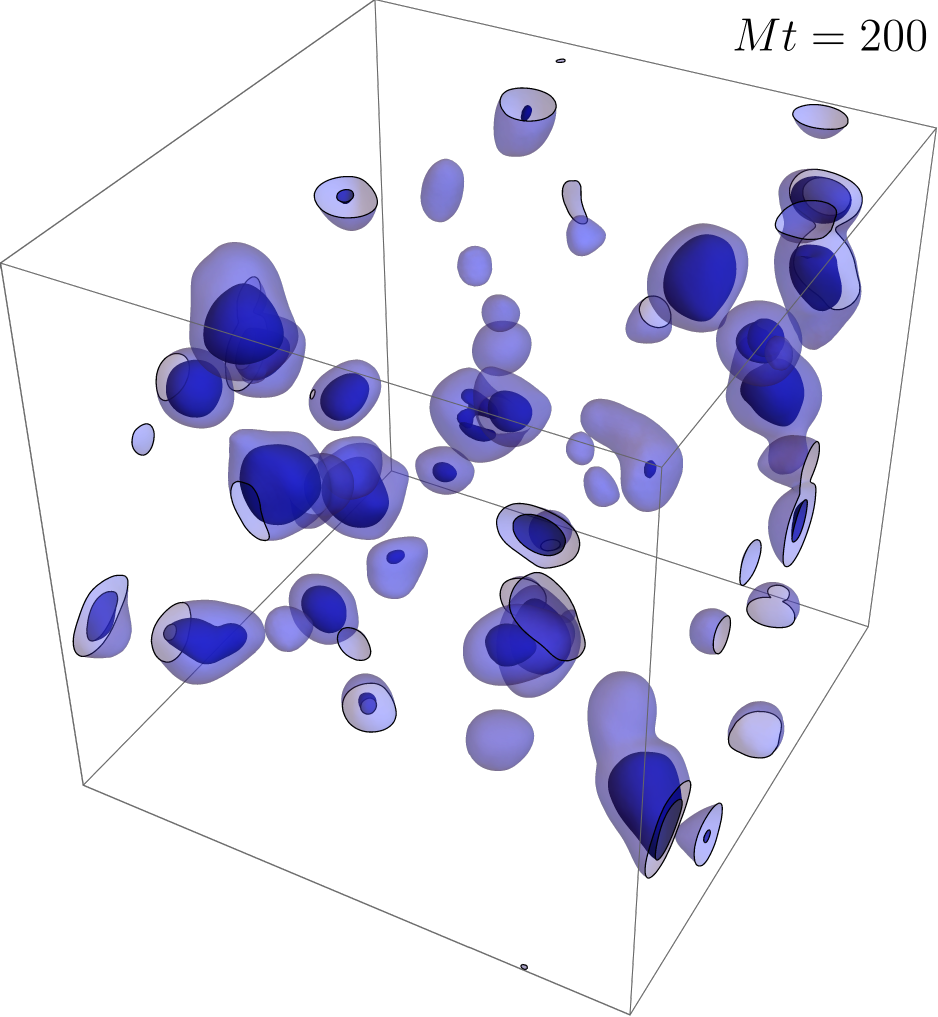}
    \end{minipage} \hfill
    \begin{minipage}{0.3\textwidth}
        \centering
        \includegraphics[width=\linewidth]{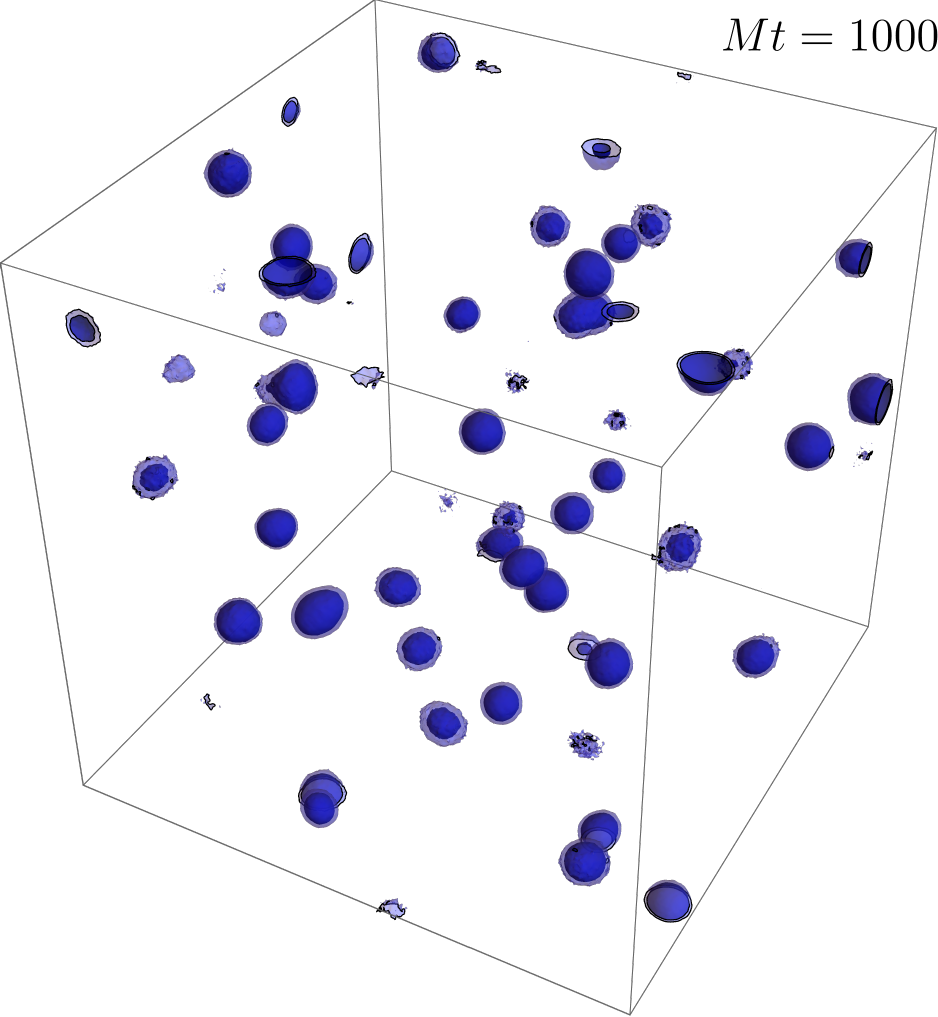}
    \end{minipage} \hfill
    \begin{minipage}{0.3\textwidth}
        \centering
        \includegraphics[width=\linewidth]{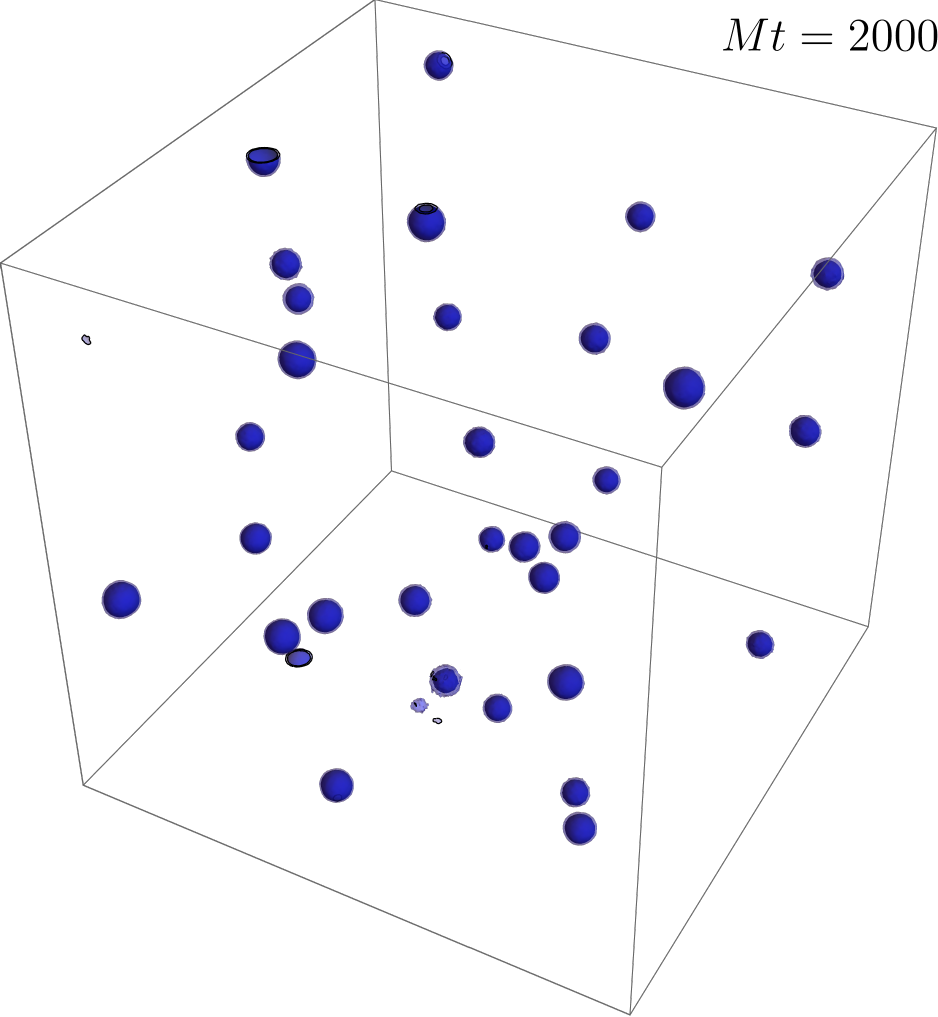}
    \end{minipage}
    \caption{Evolution of the inflaton energy overdensities during oscillon formation in Einstein-Cartan Higgs inflation (adapted from  Ref.~\cite{Piani:2025dpy}). The snapshots, extracted from 3+1 classical lattice simulations, display regions where the local energy density exceeds the average by factors of 6 (light blue) and 20 (dark blue) at various times after inflation. Early-time fragmentation generates transient overdensities, while true oscillons—localized, quasi-spherical configurations—emerge by $M\, t\simeq 500$, with $M^{-1}$ denoting the characteristic inflaton oscillation timescale.}
    \label{fig:oscillon_formation}
\end{figure}

Although the (pre)heating dynamics is often thought to be primarily dictated by the shape of the potential near its minimum, surprising effects can emerge when the potential becomes shallower than quadratic at large field values. In such cases, fragmentation can give rise to localized, long-lived structures known as oscillons~\cite{Amin:2010xe, Amin:2010dc, Amin:2011hj}: pseudostable, non-topological solitons that emerge from the nonlinear dynamics of the inflaton field, cf.~Fig.~\ref{fig:oscillon_formation}. For instance,  while the condensate does not fragment in $T$-models of inflation with $n=2$ and $\Lambda\sim M_P$,\footnote{We refer again the reader to footnote~\ref{footnote}.} long-lived oscillon configurations form efficiently for $\Lambda\ll M_P$, collectively acting as dust and dominating the energy background for many $e$-folds of post-inflationary expansion~\cite{Lozanov:2017hjm} (see also Refs.~\cite{Piani:2023aof, Piani:2025dpy}). However, for potentials steeper than quadratic ($n>2$), oscillons are generally not formed, and fragmentation leads to transient structures that decay more rapidly~\cite{Lozanov:2017hjm, Rubio:2019ypq}, giving rise to a radiation-dominated Universe.

A key consequence of fragmentation is its impact on the inflaton equation of state describing the ratio of the spatially averaged pressure $\langle p_\phi \rangle$ to the energy density $\langle \rho_\phi \rangle$. For any virialized late-time configuration [$1/2 \langle{\dot{\phi}^2}\rangle = 1/2 \langle{(\nabla\phi/a)^2}\rangle + n\langle{V}\rangle$] involving a monomial potential of the form $V(\phi)\sim \vert \phi\vert ^n$, we have 
\begin{equation}
    \langle w_\phi\rangle  \equiv \frac{\langle p_\phi \rangle}{\langle \rho_\phi \rangle} = \frac{\langle \dot{\phi}^2 / 2 - (\nabla \phi)^2 / 6 a^2 - V \rangle}{\langle \dot{\phi}^2 / 2 + (\nabla \phi)^2 / 2 a^2 + V \rangle}=\frac{1}{3}+\frac{2}{3}\frac{n-4}{(n+2)+2\langle(\nabla\phi/a)^2\rangle/\langle V\rangle}\,. 
\end{equation} 
Note that, unlike Eq.~\eqref{EoS}, this expression explicitly incorporates contributions from the gradient energy density, capturing the full nonlinear evolution of the inflaton field. Initially, when the gradients are small $\langle(\nabla\phi/a)^2\rangle \ll \langle V\rangle$, the equation of state is well approximated by the homogeneous field result in that expression, for all values of $n$. However, the late evolution is sensitive to the details of the potential~\cite{Turner:1983he, Johnson:2008se}. For $n=2$ and $\Lambda\ll M_P$, oscillons form, and $w_\phi$ remains close to zero. In contrast, for $n>2$, fragmentation leads to strong interactions among inflaton fragments, driving rapid energy redistribution regardless of the value of $\Lambda$. This process results in a radiation-dominated phase of expansion, $\omega_\phi\simeq 1/3$, for $ \langle(\nabla\phi/a)^2\rangle \gg \langle V\rangle$~\cite{Lozanov:2016hid, Lozanov:2017hjm, Bettoni:2021zhq}. 

\subsection{Stimulated boson production}
Beyond the self-interactions of the inflaton field discussed in the previous sections, non-perturbative effects can also play a crucial role in the dynamics of additional matter degrees of freedom coupled to the inflaton. In particular, bosonic fields that interact directly with the inflaton can experience efficient particle production through mechanisms such as {\it parametric}~\cite{Traschen:1990sw,Dolgov:1989us,Shtanov:1994ce,Kofman:1994rk,Kofman:1997yn} or {\it tachyonic resonance}~\cite{Felder:2000hj,Felder:2001kt,Dufaux:2006ee}, triggered by coherent oscillations of the inflaton condensate after the end of inflation. To understand the early evolution of quantum fluctuations during this stage, one can consider the linearized equation of motion for a scalar field $\varphi$, coupled to an oscillating background inflaton $\phi(t)$ in an expanding Friedmann--Lemaître--Robertson--Walker Universe. The Fourier modes of the perturbations $\delta\varphi_{\vec{k}}$ obey the general equation
\begin{equation} \label{eq:modes} 
    \delta\ddot{\varphi}_{\vec k} + 3\, H\, \delta \dot \varphi_{\vec k} + \left[\frac{k^2}{a^2} + {\cal M}_\varphi^2(\bar \phi)\right] \delta\varphi_{\vec k} = 0\,,
\end{equation}
with the effective mass term ${\cal M}_\varphi^2$ depending on the background inflaton field $\bar{\phi}(t)$. This general form serves as a starting point for analyzing a wide variety of (pre)heating scenarios. In particular, in the presence of three-legged vertex scalar interactions $\mu\, \phi\, |\varphi|^2$, the daughter particles $\varphi$ acquire an effective mass ${\cal M}_\varphi^2 = \mu\, \phi(t)$ due to the oscillating inflaton condensate. The trilinear vertex therefore results in a tachyonic mass of $\varphi$, whenever $\phi(t)<0$. Consequently, the modes satisfying $k^2/a^2 < \mu\, |\phi|$ will be exponentially amplified during a portion of each half-period of the inflaton oscillations. This trilinear vertex therefore gives rise to both parametric resonance~\cite{Shtanov:1994ce, Kaiser:1995fb, Kofman:1994rk, Kofman:1997yn} and {\it tachyonic (pre)heating}~\cite{Felder:2000hj, Felder:2001kt}, leading to what is known as tachyonic resonance~\cite{Greene:1997ge, Dufaux:2006ee, Abolhasani:2009nb}. This results in very efficient production of the daughter particles, which is not captured by the perturbative calculations. If the amplitude $\Phi$ of inflaton oscillations satisfies $\mu\, \Phi > m_\phi^2$ in the trilinear case, high-order Feynman diagrams give comparable predictions to the lowest-order ones and the problem has to be approached non-perturbatively. Similarly to the case of the three-legged vertex, in the four-legged vertex it is also possible to realize the effect of parametric resonance. In particular, if the effective mass of the daughter field ${\cal M}_\varphi^2 = 2\, g\, \phi^2(t)$ is much greater than that of the inflaton inflaton, $\sqrt{2\, g}\, \Phi/m_\phi \gg 1$, the corresponding frequency in Eq.~\eqref{eq:modes} becomes much higher than that of the inflaton, leading to a broad parametric resonance regime whose early stages can be studied using a WKB approximation~\cite{Kofman:1994rk, Kofman:1997yn}. In this scenario, the resonant production of the particles occurs over a broad range of $k$, and the (re)heating becomes extremely efficient.

The homogeneous oscillations of the inflaton leading to rapid growth of spatially varying perturbations via parametric or tachyonic resonance cannot proceed forever. As the occupation numbers of the produced particles grow, their collective energy density becomes significant enough to influence the dynamics of the inflaton field~\cite{Khlebnikov:1996wr, Khlebnikov:1996zt, Prokopec:1996rr}. This backreaction effect not only alters the effective potential of the inflaton but also modifies the resonance structure inherent in parametric amplification, leading to the shift or suppression of resonance bands. The simplest way to take into account the backreaction of the amplified quantum fluctuations is to use the Hartree approximation~\cite{Kofman:1994rk, Kofman:1997yn}, in which different modes and fields evolve independently (uncorrelated in time). However, as the number of particles increases, the mean-field/Hartree approximation stops being a good description, and coupling between different Fourier modes becomes important,  necessitating a fully non-linear treatment to accurately capture the evolution of the system. Moreover, the produced particles interact among themselves and with the inflaton field through rescattering processes. These interactions lead to a redistribution of energy among different field modes, contributing to the fragmentation of the inflaton condensate and the approach toward thermal equilibrium.

\subsection{Multifield blocking effects}
The picture presented in the previous section may change significantly in the presence of multiple scalar fields $\phi_I$ that contribute to inflation. In such multifield setups, each inflaton component typically oscillates with a different mass and amplitude, leading to a superposition of oscillatory modes. If all the fields are coupled to a common matter species---for instance, through interaction terms like $\sum_I g_I^2\, \phi_I^2\, \varphi^2$---the resulting effective mass of the daughter field $\varphi$ becomes a time-dependent function with contributions from all $\phi_I(t)$ components. Due to the differing frequencies, phases, and decay rates of these inflaton oscillations, the combined evolution often lacks the regularity required for efficient resonance. This loss of coherence between the inflaton fields---sometimes referred to as \textit{de-phasing}---can ``block'' the amplification of the daughter field's fluctuations by smoothing out the zero crossings of its effective mass \cite{Battefeld:2008bu,Battefeld:2009xw}. As a result, sharp adiabaticity violations necessary for triggering explosive particle production may be suppressed. This phenomenon reduces the efficiency of (pre)heating, delaying energy transfer to the radiation bath, and potentially prolonging the (re)heating phase.

Despite the aforementioned suppression, multifield models are not necessarily insensitive to non-perturbative processes. The structure of the couplings plays a critical role. For example, introducing trilinear couplings or allowing negative couplings for certain inflaton components can restore or even enhance the resonant instabilities, bypassing the blocking effects associated with simple quadratic couplings \cite{Dufaux:2006ee}.

\subsection{The depleting role of fermions}
Fermions can also be produced non-perturbatively from the inflaton field during (pre)heating, but their production is severely constrained by the Pauli exclusion principle, which limits their number and reduces their direct impact on the inflaton field~\cite{Greene:1998nh,Baacke:1998di,Greene:2000ew, Peloso:2000hy}. However, in models where resonantly produced bosons decay into fermions, these fermions act as a ``spillway'' draining energy from the system and alleviating backreaction on the inflaton. This delays the onset of parametric resonance effects, providing a more controlled and gradual evolution of the system~\cite{Garcia-Bellido:2008ycs, Rubio:2015zia, Repond:2016sol}. Unlike traditional (pre)heating models, where backreaction typically halts further energy release from the inflaton, this \textit{combined (pre)heating} mechanism facilitates up to a four-order magnitude increase in energy dissipation, with the remaining inflaton energy reduced to as little as $0.01\%$~\cite{Fan:2021otj, Mansfield:2023sqp}.

The equation of state of the Universe during (pre)heating is also influenced by the presence of fermions. In conventional (pre)heating models, the energy density evolves from a mixture of matter and radiation. However, in combined (pre)heating, the rapid decay of fermions accelerates the transition to a radiation-dominated state, making the evolution of the Universe faster and more efficient. As fermions act as radiation-like components, they contribute to the overall energy budget, leading to a quicker approach to the characteristic value $w = 1/3$ of radiation domination~\cite{Mansfield:2023sqp}.

\subsection{Gravitational (pre)heating}
The non-perturbative transfer of energy from the inflaton to matter fields can also occur via gravitational interactions, provided that a scalar spectator field $\varphi$---either from the SM or beyond---is non-minimally coupled to gravity, such that its initial dynamics is governed by the effective equation of motion \eqref{eq:modes} with ${\cal M}^2_\varphi = \xi\, R$, where $\xi$ is the non-minimal coupling to gravity. A well-motivated setting for this mechanism arises in quintessential inflation scenarios, where the Universe undergoes a kination phase ($w_\phi = 1$) following the end of inflation. During this transition, the Ricci scalar, $R = 6\, (\dot{H} + 2\, H^2) = 3\, (1 - 3\, w_\phi)\, H^2$, becomes negative and induces an effective tachyonic mass ${\cal M}^2_\varphi < 0$ for the spectator field $\varphi$.  This results in a burst of nonadiabatic particle production via tachyonic instability, leading to the exponential amplification of quantum fluctuations~\cite{Bettoni:2019dcw, Bettoni:2021zhq, Laverda:2023uqv}. As first pointed out in Refs.~\cite{Felder:2000hj, Felder:2001kt, Bettoni:2019dcw}, any perturbative picture of a homogeneously oscillating scalar field fails to capture the full dynamics of spontaneous symmetry breaking. In particular, this simplified view also overlooks crucial effects such as the formation of topological defects~\cite{Bettoni:2018pbl,Bettoni:2019dcw,Bettoni:2021zhq}, which can have significant observational consequences, including the generation of a stochastic GW background~\cite{Bettoni:2018pbl, Bettoni:2024ixe} and baryogenesis~\cite{Bettoni:2018utf, Chen:2025awt}.

As shown in Refs.~\cite{Laverda:2024qjt, Laverda:2025pmg}, the explosive particle production during a Hubble-induced tachyonic phase allows the Higgs field itself to be responsible for the onset of the hot Big Bang era, allowing for (re)heating temperatures in the range $10^{-2}-10^9$ GeV and opening the gate to implement potential electroweak baryogenesis mechanisms~\cite{Shaposhnikov:1987tw, Wagner:2023vqw}. The viable parameter space, particularly the relationship between the Higgs mass, the inflationary scale, and the top-quark mass, is constrained by both vacuum stability requirements and the need to achieve successful (re)heating before BBN~\cite{Laverda:2024qjt, Laverda:2025pmg}.  In general, these constraints favor a lower top-quark pole mass, in agreement with current measurements~\cite{CMS:2023ebf, Myllymaki:2024uje}.

\subsection{Towards thermalization}
After the initial stages of (pre)heating, where non-perturbative processes lead to explosive particle production, the Universe transitions toward thermalization—a phase characterized by the establishment of thermal equilibrium among the produced particles. This process is intricate, involving various stages and mechanisms that collectively drive the system from a highly non-equilibrium state to one of local thermal equilibrium. 

Immediately after fragmentation, the Universe is populated by a dense assembly of interacting fields with large occupation numbers, rendering a classical field theory description appropriate. The energy spectra of the particles produced during (pre)heating are usually highly non-thermal, exhibiting specific peaks corresponding to resonant modes. As interactions proceed, processes such as scattering and decay lead to a redistribution of energy and the establishment of a turbulent dynamics~\cite{Khlebnikov:1996mc, Greene:1997fu, Micha:2002ey, Micha:2004bv}, where energy is transferred across different scales, a phenomenon known as an energy cascade. This turbulent behavior is analogous to wave turbulence observed in other physical systems and plays a pivotal role in the redistribution of energy among the modes of the fields. The particle spectra during a turbulent stage exhibit generically a momentum dependence $n_k\sim k^{-3/2}$ and a self-similar evolution, characterized by specific scaling behaviors in the distribution functions of the fields. For instance, in the case of a $\lambda\, \phi^4$ model, one has~\cite{Micha:2002ey, Micha:2004bv}
\begin{equation}
    n_k(\tau) = \tau^{-q}\, n_0(k\, \tau^{-p})\,,
\end{equation}
with $\tau = \eta / \eta_c$ a rescaled conformal time, $n_0(k)$ the distribution function at a later time $\eta_c$ within the self-similar regime and $q \sim 3.5\, p$ and $p \sim 1/5$~\cite{Micha:2002ey, Micha:2004bv}. The presence of such spectra suggests that the system is undergoing a universal process of energy transfer from small- to large-momentum scales, regardless of the specific details of the initial conditions. Such a self-similar dynamic, characteristic of turbulent Kolmogorov spectra, is crucial for understanding how the system approaches thermal equilibrium over time. 

Before reaching complete thermalization, the system may enter a state known as {\it prethermalization} where certain macroscopic quantities, such as the equation of state, approach values close to those of thermal equilibrium, even if this is not yet locally established~\cite{Felder:2000hr, Berges:2004ce, Podolsky:2005bw, Lozanov:2016hid}. Specifically, the global equation of state parameter tends toward $1/3$, signaling a radiation-dominated Universe. However, true local thermal equilibrium requires both kinetic equilibrium (a uniform distribution of energy and momentum among particles) and chemical equilibrium, with number-changing processes establishing the equilibrium abundances of different species. Although prethermalization can occur relatively quickly, full thermalization is a more gradual process, governed by the rates of these microscopic interactions. Capturing this final evolution presents a significant challenge for classical lattice simulations, as they are fundamentally limited by the well-known Rayleigh-Jeans divergence~\cite{Berges:2013lsa}. In particular, in classical field theory, all modes are thermally occupied according to the Rayleigh-Jeans law, leading to an ultraviolet divergence in energy density at finite temperature. In the continuum limit, this results in an unphysical scenario where the temperature approaches zero, while on a lattice, it remains sensitive to the imposed discretization scale. As a result, while lattice simulations effectively describe early nonlinear dynamics—such as self-resonance, fragmentation, and turbulent energy cascades—they fail to accurately capture the true quantum thermalization process. To properly account for the final approach to equilibrium, more standard techniques, such as quantum BEQs or statistical field theory approaches, are often required~\cite{Micha:2002ey, Micha:2004bv, Kurkela:2011ti}.

\section{Rare analytic birds} \label{sec:nolattice}
As discussed in the previous sections and summarized in Fig.~\ref{fig:(re)heating-stages} and Table~\ref{tab:(pre)heating_misconceptions}, the treatment of (re)heating is often far more intricate than typically assumed within the particle physics community, frequently requiring advanced numerical techniques that go far beyond the simplistic Boltzmann approach. This naturally raises the question: must one always resort to daunting tools like lattice simulations, spending countless hours poring over evolving field configurations? While the general answer is regrettably affirmative, there do exist exceptional scenarios where analytic control can still be partially retained, allowing for meaningful progress without the full machinery of non-perturbative methods. In this section, we highlight some of these \textit{rara avis}.
\begin{table}
    \centering
    \renewcommand{\arraystretch}{1.4}
    \setlength{\tabcolsep}{4pt}
    \footnotesize
    \begin{tabular}{
        >{\columncolor{blue!10}\raggedright\arraybackslash}p{2.5cm}
        >{\raggedright\arraybackslash}p{4cm}
        >{\raggedright\arraybackslash}p{4cm}
        >{\raggedright\arraybackslash}p{4cm}
        }
        \toprule
        \rowcolor{blue!20}
        \multicolumn{4}{c}{\textbf{I. Foundational Assumptions}} \\
        \midrule
        \makecell{\textbf{Topic}} & 
        \makecell{\textbf{Typically assumed}} & 
        \makecell{\textbf{Commonly happening}} & 
        \makecell{\textbf{Conceptual Implications}} \\
        \midrule
        \textbf{Fields} & Treated as particle gases. &  Behave as classical waves for large occupation numb. & Standard particle-based tools become unreliable. \\
        \textbf{Quantum stat.} & Often ignored. Fermions treated like bosons. & Bose effects enhance growth Pauli exclusion limits it. & Quantum statistics shapes particle production. \\
        \rowcolor{blue!20}
        \multicolumn{4}{c}{\textbf{II. Inflaton and Energy Transfer}} \\
        \midrule
        \textbf{Inflaton} & Smooth and oscillating condensate. & Breaks into inhomogeneous lumps/solitons/defects. & Full field dynamics needed.  Lattice methods required. \\
        \textbf{Energy Transf.} & Slow, perturbative decays. Expansion neglected. & Fast, explosive  instabilities.   Redshift affects resonances.  & Nontrivial energy loss.  Lattice methods required.   \\
        \textbf{Backreaction} & Homogeneity unaffected. Nonlinearities ignored. & Breaking of homogeneity.\quad \quad \quad Rescatterings \& turbulence. & Fully coupled dynamics. Lattice methods required. 
        \\
        \rowcolor{blue!20}
        \multicolumn{4}{c}{\textbf{III. Spectrum, Timescales and  Thermalization}} \\
        \midrule
        \textbf{Spectrum} & Narrow and peaked at production & Broad with distinct bands. & Spectral structure controls the dynamics. \\
        \textbf{Timescale} & Radiation domination onset takes many oscillations. & (Re)heating can complete in just a few oscillations. & Transition may be nearly instantaneous. \\
        \textbf{Thermalization} & Instant once energy is transferred. & Delayed by chaotic field behavior. & Needs modeling of prethermal stages. \\
        \textbf{Final State} & Equilibrium bath with clear temperature. & Chaotic fields far from equilibrium. & Must distinguish early and final temperatures. \\
        \bottomrule
    \end{tabular}
    \caption{Comparison between the Boltzmann ((re)heating) and non-perturbative ((pre)heating) approaches.}
    \label{tab:(pre)heating_misconceptions}
\end{table}

\subsection{Starobinsky scenario: a perturbative take}
Among the rare instances where a perturbative treatment of (re)heating remains reliable, the Starobinsky model of inflation stands out as a prime example. The standard action of this scenario takes the form~\cite{Starobinsky:1980te, Starobinsky:1981vz, Starobinsky:1983zz, Kofman:1985aw, Bernal:2020qyu}
\begin{equation} \label{eq:S_staro}
    S = \frac{M_P^2}{2} \int d^4x\, \sqrt{-\tilde g} \left[\tilde R + \frac{\tilde R^2}{6 M^2}\right] + S_M(\tilde g_{\mu\nu},\varphi)\,,
\end{equation}
with $S_M$ a matter action containing all SM model fields, and in particular, the Higgs doublet $\varphi$ minimally coupled to gravity. The presence of the term $\tilde R^2$ in Eq.~\eqref{eq:S_staro} allows an inflationary state able to generate the observed amount of primordial density perturbations ${\cal P} \simeq 2.1\times 10^{-9}$~\cite{Planck:2018jri} for suitable values of the mass parameter $M$, namely $M\simeq 1.3\times  10^{-5} \left(54/N_*\right) M_P $. Here, $N_*$ corresponds to the number of $e$-folds of inflation needed to solve the flatness and horizon problems, which depends itself on the whole post-inflationary history and, in particular, on the details of the heating stage, which we now describe.

Unlike other approaches that require introducing couplings between the inflaton and matter fields by hand, the Starobinsky scenario offers a predictive and universal framework where all interactions arise unambiguously from gravitational couplings. In particular, once the theory is defined in the frame~\eqref{eq:S_staro} with a non-minimally coupled matter sector, no further assumptions are needed: no ad hoc interaction terms, arbitrary branching ratios, or speculative particle content. To see this explicitly, one can perform a Weyl transformation $\tilde g_{\mu\nu} = \Omega^2(\phi)\, g_{\mu\nu}$ with conformal factor $ \Omega^2 = \exp(\sqrt{2/3}\, \phi/M_P)$, such that the action is recast as~\cite{Stelle:1977ry, Whitt:1984pd, Mukhanov:1989rq}
\begin{equation} \label{eq:EFaction}
    S =  \int d^4x \sqrt{-g} \left[\frac{M_P^2}{2} R - \frac12\, g^{\mu\nu}\, \partial_\mu\phi\, \partial_\nu\phi - \frac34\, M_P^2\, M^2 \left(1 - e^{-\sqrt{\frac23} \frac{\phi}{M_P}}\right)^2\right] + S_M(\Omega^2(\phi) g_{\mu\nu},\varphi)\,,
\end{equation}
with the so-called \textit{scalaron} field $\phi$ playing the role of the inflaton field. In this Einstein or scalaron frame, the action for the nonconformally coupled Higgs boson inherits universal couplings to the scalaron through the conformal rescaling of the metric. Crucially, the rescaled field-dependent mass term in this expression never vanishes or becomes negative during the post-inflationary evolution, preventing the emergence of tachyonic instabilities or resonance bands that would otherwise trigger non-perturbative particle production. In addition, the formation of massive oscillons, typically associated with potentials that away from the minimum are shallower than quadratic, does not take place in this setting since $\Lambda \sim M_P$~\cite{Takeda:2014qma}.

The absence of non-perturbative phenomena greatly simplifies the post-inflationary dynamics. The scalaron decays perturbatively into the Higgs field via gravitationally induced interactions, whose strength is fixed by the model and not subject to tuning.\footnote{We note that, although absent at tree level, the anomalous decay of the scalaron field into gauge bosons is possible at 1-loop. In particular, the breaking of scale symmetry during the regularization process translates into an induced breaking of the gauge conformal symmetry~\cite{Watanabe:2010vy}.} The resulting decay width can be computed analytically via the Bogoliubov's method in the original frame~\eqref{eq:S_staro}~\cite{Starobinsky:1980te, Starobinsky:1981vz} or through standard perturbative techniques in the scalaron frame~\eqref{eq:EFaction}~\cite{Watanabe:2006ku, Watanabe:2010vy}, obtaining in both cases~\cite{Rudenok:2014daa, Ema:2016hlw}
\begin{equation} \label{Gamma_varphi}
    \Gp \simeq
    \frac{1}{24 \pi} \frac{M^3}{M_P^2} + \mathcal{O}\left(\frac{m_h}{M}\right)^2 \simeq 2.9 \times 10^{-17}\, M_P\,,
\end{equation} 
where we have taken into account that the Higgs mass $m_h$ is much smaller than the scalaron mass, neglecting therefore phase-space suppression factors. Similarly, assuming instantaneous thermalization of the decay products,\footnote{Strictly speaking, the decay products are initially distributed with smaller occupation numbers and harder momenta, $\langle k \rangle \simeq M$, compared to those in a thermal distribution~\cite{Harigaya:2013vwa, Ellis:2015jpg, Garcia:2018wtq}. As a result, SM particles become approximately thermalized only at a later stage, characterized by the temperature $T_{\rm th} \simeq \alpha^{4/5} M \left(\Gp\, M_P^2/M^3 \right)^{2/5}$, with $\alpha$ being the fine structure constant of the gauge interaction~\cite{Harigaya:2013vwa}. For a minimally coupled Higgs and $\alpha \simeq 10^{-2}$, this corresponds to $T_{\rm th} \simeq 10^{11}$~GeV; about one order of magnitude below $\Tmax$. Since $\Tmax$ depends more weakly on $\Gp$ than does $T_{\rm th}$, the thermalization temperature can exceed the maximum temperature, $T_{\rm th} > \Tmax$, if $\Gp > 2 \times 10^{-9}M_P$, which justifies the assumption of instantaneous thermalization. On the other hand, if $\Tmax \gtrsim T_{\rm th}$, then $\Tmax$ should be interpreted as an effective measure of the maximal radiation density as defined by Eq.~\eqref{eq:Tmax}.} the maximal effective radiation temperature of the associated SM plasma can be estimated as
\begin{equation} \label{eq:Tmax}
	\Tmax = \left(\frac{30\, \rR(\Tmax)}{\pi^2\,\gs} \right)^{1/4}
	\simeq \left(\frac{5\sqrt{3}}{\pi^2\,\gs} \sqrt{\,V}\, M_P\, \Gp\right)^{1/4} \simeq 2\times 10^{12}~\text{GeV}\,.
\end{equation}
The temperature of the SM plasma within this period is therefore in the range $\Trh \lesssim T \lesssim \Tmax$, with $\Trh \simeq \gs^{-1/4} \sqrt{\Gp\, M_P} \simeq 4.2 \times 10^9$~GeV the standard (re)heating temperature, defined approximately by the moment at which the total decay width of the scalaron into the SM components equals the Hubble rate, $\Gp = 3\, H$~\cite{Gorbunov:2010bn, Gorbunov:2012ns}.

\subsection{Non-oscillatory potentials: an analytical non-perturbative approach}
In contrast to standard (pre)heating scenarios that require prolonged oscillations of the inflaton and lattice simulations to follow nonlinear evolution, a particularly elegant and analytically tractable mechanism known as \textit{instant (pre)heating} emerges in non-oscillatory models~\cite{Felder:1998vq,Felder:1999pv}. In these scenarios, the inflaton field does not oscillate after inflation but rolls monotonically, enabling non-perturbative particle production through a single, sharp violation of adiabaticity. This one-time event allows the entire mechanism to be described analytically, bypassing the need for nonlinear numerical tools. For this to happen, the effective mass term ${\cal M}_\varphi^2$ for the matter field $\varphi$ should be $i)$ large enough during inflation as to retain the single-field inflationary dynamics, $ii)$ vary rapidly at the end of inflation to heat the Universe via adiabaticity violations, and, depending on the model, $iii)$ decrease monotonically with time in order to avoid strong backreaction effects at large $\phi$ values. A simple choice satisfying all these criteria is, for instance,\footnote{Alternative options that share the characteristics described in $i)$, $ii)$, and $iii)$ could be used without significantly modifying the conclusions below. For example, one could introduce a parameter $\phi_l$ encoding the timing of the transition simply by replacing $\phi$ by $\phi - \phi_l$ or one could consider smoothing the transition at $\phi = 0$ by using some interpolation function; see, e.g. Ref.~\cite{Rubio:2017gty}. Note also that condition $iii)$ could be relaxed if the daughter field $\varphi$ is coupled to lighter fermionic degrees of freedom. Their inclusion significantly increases the efficiency of (re)heating, draining energy from the $\varphi$ sector before the backreaction on $\phi$ becomes important~\cite{Felder:1998vq,Felder:1999pv}. In the most efficient scenarios, (re)heating can become nearly instantaneous.}
\begin{equation} \label{mass}
    {\cal M}_\varphi^2(\phi) =
    \begin{dcases} 
        g^2\, \phi^2 & \quad\text{ for }\phi\leq 0\,, \\
        \tilde m^2_\varphi & \quad\text{ for }\phi >  0\,,
   \end{dcases}
\end{equation}
with $\tilde m_\varphi$ a constant. This a priori unconventional behavior is expected in models where quintessential inflation is associated with the emergence of quantum scale symmetry in the vicinity of UV and IR fixed points~\cite{Wetterich:2014gaa, Rubio:2017gty, Bettoni:2021qfs}. In this type of variable gravity scenario~\cite{Wetterich:2013jsa}, rapid variations of ${\cal M}_\varphi^2(\phi)$ are only expected to occur in a crossover regime where the dimensionless couplings and mass ratios of matter fields evolve from their UV to IR values. If the field $\varphi$ is identified with the SM Higgs, the decoupling from $\phi$ in Eq.~\eqref{mass} at late times encodes the approach to the SM IR fixed point, as required by the constraints on the variation of the Fermi to Planck mass ratio since nucleosynthesis~\cite{Uzan:2010pm, Wetterich:2003qb}.

The form of Eq.~\eqref{mass} ensures that ${\cal M}_\varphi^2$ changes non-adiabatically only once near $\phi = 0$, triggering an instantaneous burst of particle production. At small $k$ values, the violation of the adiabaticity condition can be safely approximated by $|\dot {\cal M}_{\varphi}| \gtrsim {\cal M}^2_{\varphi}$ or equivalently by $g^2\, \phi^2 \lesssim g\, |\dot\phi_0|$, with $\vert \dot \phi_0\vert $ the inflaton velocity at zero crossing. Solving this expression for $\phi$, we observe that particle production takes place in a very narrow interval $\Delta \phi \sim (|\dot\phi_0|/ g)^{1/2}$ around $\phi = 0$, the production being essentially instantaneous for sufficiently large couplings, $\Delta t \sim  \Phi/|\dot\phi_0| \sim (g\, |\dot\phi_0|)^{-1/2}$. The typical momentum of the created particles follows directly from the uncertainty principle, $\Delta k \sim (\Delta t)^{-1} \sim (g\, |\dot\phi_0|)^{1/2}$ and coincides with the one obtained by properly solving the mode equation~\eqref{eq:modes} in the WKB approximation. Indeed, as shown explicitly in the seminal paper~\cite{Kofman:1997yn}, the occupation number of $\varphi$ particles after a single zero crossing is given by
\begin{equation} \label{number}
    n^{\rm kin}_k = \exp\left(- \frac{\pi\, k^2}{g\, |\dot\phi_0|}\right).   
\end{equation}
Assuming the decay products to be ultra-relativistic ($g\, |\dot\phi_0| \gg \tilde m_\varphi$), this corresponds to an instantaneous generation of the radiation energy density $\rho^{\rm kin}_R \simeq g^2/(4 \pi ^4)\, \vert \dot \phi_0\vert^2$, which redshifts as $\rho_R \propto a^{-4}$. Since no further non-adiabatic transitions occur, there is no subsequent resonance, and the entire process is completed in a single event, in stark contrast with oscillatory models where repeated zero-crossings lead to parametric resonance and require careful numerical treatment. On top of that, the results are completely independent of the particle spin, allowing us to extend the above estimates to fermionic species. We emphasize that this is not the case in oscillatory scenarios, since the adiabaticity condition is violated periodically, leading to bosonic enhancement effects~\cite{Kofman:1994rk, Kofman:1997yn}.

Using the scaling of the different energy components during the kination epoch ($\rho_\phi(a) \propto a^{-6}$ and $\rho_R(a) \propto a^{-4}$) together with entropy conservation, one can easily determine the (re)heating temperature at which the energy density of the created particles equals that of the inflaton field ($\rho_R(\Trh) = \rho_\phi(\Trh)$), namely 
\begin{equation} \label{Treh0}
    \Trh = \left(\frac{\gss(\Tmax)}{\gss(\Trh)}\right)^{1/3} \Theta^{1/2}\, \Tmax\,,
\end{equation}
with $\gss(\Tmax)$ and $\gss(\Trh)$ the entropic degrees of freedom at the corresponding temperature scales, $\Tmax^4 \equiv 30\,\rho_{\rm R}^{\rm kin}/({\pi^2\,\gs(\Tmax)})$ the maximum temperature of the created particles at the onset of kination under the assumption of instantaneous thermalization, and 
\begin{equation}
    \Theta \equiv \frac{\rho_R^{\rm kin}}{\rho_{\phi}^{\rm kin}} \simeq 2 \times 10^{-8} \left(\frac{g}{0.02}\right)^2 \left(\frac{10^{11} \, {\rm GeV}}{H_{\rm kin}}\right)^2\left(\frac{\vert \dot \phi_0\vert }{10^{-8} M_P^2}\right)^2
\end{equation} 
the so-called \textit{heating efficiency} encoding the efficiency of non-perturbative particle production~\cite{Rubio:2017gty, Bernal:2020bfj}. We observe then that the smaller the heating efficiency, the longer the kination epoch, and the larger the difference between the maximal radiation temperature and the proper (re)heating temperature. In particular, accounting for all SM degrees of freedom at temperatures higher than the top-quark mass ($\gs(\Tmax) = \gss(\Tmax) = \gss(\Trh) = 106.75$), the upper limit on the inflationary scale $H_I < 2.5\times10^{-5}\, M_P$~\cite{Planck:2018jri} implies a bound $\Tmax \leq 6 \times 10^{11}~\text{GeV} \times \Theta^{1/4}$.

\section{Conclusions} \label{sec:concl}
As particle physics and cosmology continue to further intertwine, the (re)heating epoch stands out as a pivotal bridge between high-energy inflationary dynamics and the thermal history of the observable Universe. A key takeaway from this review is that, despite its intuitive appeal and analytical tractability, the conventional Boltzmann approach—commonly used within the particle physics community—fails to capture the essential nonlinear and non-perturbative phenomena that dominate the early stages of (re)heating. These include explosive particle production via parametric and tachyonic resonance, condensate fragmentation, and turbulent energy cascades; processes that are not just theoretical curiosities but critical components of the post-inflationary dynamics.

Although significant uncertainties concerning the detailed evolution from the end of inflation to radiation domination remain, decades of theoretical progress and increasingly sophisticated computational tools, ranging from lattice simulations to semi-analytic techniques, have brought us closer to a coherent and predictive framework. One of the aims of this review has been to demystify these developments, highlight the physical processes at play, and offer to the particle physics community a practical entry point into this rich area of early-Universe dynamics. Moreover, the dynamics of (pre)heating is not merely of academic interest: they can leave observable imprints. The violent amplification of field fluctuations can source stochastic gravitational waves~\cite{Caprini:2018mtu}, contribute to baryogenesis~\cite{Kolb:1998he, Giudice:1999fb, Krauss:1999ng, Garcia-Bellido:1999xos, Copeland:2001qw, Garcia-Bellido:2003wva, Tranberg:2006dg}, affect the primordial power spectrum~\cite{Bassett:1998wg, Finelli:1998bu, Chambers:2007se, Bond:2009xx, Imrith:2019njf, Giblin:2019nuv, Adshead:2023mvt}, and lead to the formation of topological defects such as cosmic strings or domain walls~\cite{Kasuya:1998aq, Rajantie:2000fd, Felder:2000hj, Copeland:2002ku, Dufaux:2010cf, Bettoni:2018pbl, Bettoni:2019dcw, Bettoni:2021zhq}. These cosmological relics offer a unique window into the Universe's earliest moments and provide strong motivation for connecting early-Universe dynamics with upcoming observational probes.

In summary, (re)heating is not a footnote to inflation, nor a simple perturbative afterthought. It is a rich and intricate phase that demands both conceptual clarity and methodological care. As cosmology becomes an increasingly precise science grounded in data, faithful modeling of this transitional epoch will be essential not only for consistency but also for unlocking new insights into high-energy physics and the fundamental structure of our Universe.
 
\acknowledgments
BB and NB, known to have previously relied a bit too faithfully on the Boltzmann approach, hereby publicly commit to respecting its limitations; at least until the next draft. NB received funding from the grant PID2023-151418NB-I00 funded by MCIU/AEI/10.13039/ 501100011033/FEDER, UE. JR is supported by a Ramón y Cajal contract of the Spanish Ministry of Science and Innovation with Ref.~RYC2020-028870-I. This research was further supported by the project PID2022-139841NB-I00 of MICIU/AEI/10.13039/501100011033 and FEDER, UE. JR thanks Matteo Piani for valuable assistance with oscillon snapshots.

\bibliographystyle{JHEP}
\bibliography{biblio}

\providecommand{\href}[2]{#2}\begingroup\raggedright\begin{thebibliography}{100}

\bibitem{Kaneta:2022gug}
K.~Kaneta, S.M.~Lee and K.-y.~Oda, \emph{{Boltzmann or Bogoliubov? Approaches
  compared in gravitational particle production}},
  \href{https://doi.org/10.1088/1475-7516/2022/09/018}{\emph{JCAP} {\bfseries
  09} (2022) 018} [\href{https://arxiv.org/abs/2206.10929}{{\ttfamily
  2206.10929}}].

\bibitem{Bassett:2005xm}
B.A.~Bassett, S.~Tsujikawa and D.~Wands, \emph{{Inflation dynamics and
  reheating}}, \href{https://doi.org/10.1103/RevModPhys.78.537}{\emph{Rev. Mod.
  Phys.} {\bfseries 78} (2006) 537}
  [\href{https://arxiv.org/abs/astro-ph/0507632}{{\ttfamily
  astro-ph/0507632}}].

\bibitem{Allahverdi:2010xz}
R.~Allahverdi, R.~Brandenberger, F.-Y.~Cyr-Racine and A.~Mazumdar,
  \emph{{Reheating in Inflationary Cosmology: Theory and Applications}},
  \href{https://doi.org/10.1146/annurev.nucl.012809.104511}{\emph{Ann. Rev.
  Nucl. Part. Sci.} {\bfseries 60} (2010) 27}
  [\href{https://arxiv.org/abs/1001.2600}{{\ttfamily 1001.2600}}].

\bibitem{Amin:2014eta}
M.A.~Amin, M.P.~Hertzberg, D.I.~Kaiser and J.~Karouby, \emph{{Nonperturbative
  Dynamics Of Reheating After Inflation: A Review}},
  \href{https://doi.org/10.1142/S0218271815300037}{\emph{Int. J. Mod. Phys. D}
  {\bfseries 24} (2014) 1530003}
  [\href{https://arxiv.org/abs/1410.3808}{{\ttfamily 1410.3808}}].

\bibitem{Lozanov:2019jxc}
K.D.~Lozanov, \emph{{Lectures on Reheating after Inflation}},
  \href{https://arxiv.org/abs/1907.04402}{{\ttfamily 1907.04402}}.

\bibitem{Lozanov:2020zmy}
K.~Lozanov, \emph{{Reheating After Inflation}}, SpringerBriefs in Physics,
  Springer (9, 2020),
  \href{https://doi.org/10.1007/978-3-030-56810-8}{10.1007/978-3-030-56810-8}.

\bibitem{Kolb:1990vq}
E.W.~Kolb, \emph{{The Early Universe}}, vol.~69, Taylor and Francis (5, 2019),
  \href{https://doi.org/10.1201/9780429492860}{10.1201/9780429492860}.

\bibitem{Albrecht:1982mp}
A.~Albrecht, P.J.~Steinhardt, M.S.~Turner and F.~Wilczek, \emph{{Reheating an
  Inflationary Universe}},
  \href{https://doi.org/10.1103/PhysRevLett.48.1437}{\emph{Phys. Rev. Lett.}
  {\bfseries 48} (1982) 1437}.

\bibitem{Giudice:2003jh}
G.F.~Giudice, A.~Notari, M.~Raidal, A.~Riotto and A.~Strumia, \emph{{Towards a
  complete theory of thermal leptogenesis in the SM and MSSM}},
  \href{https://doi.org/10.1016/j.nuclphysb.2004.02.019}{\emph{Nucl. Phys. B}
  {\bfseries 685} (2004) 89}
  [\href{https://arxiv.org/abs/hep-ph/0310123}{{\ttfamily hep-ph/0310123}}].

\bibitem{Buchmuller:2004nz}
W.~Buchmuller, P.~Di~Bari and M.~Plumacher, \emph{{Leptogenesis for
  pedestrians}}, \href{https://doi.org/10.1016/j.aop.2004.02.003}{\emph{Annals
  Phys.} {\bfseries 315} (2005) 305}
  [\href{https://arxiv.org/abs/hep-ph/0401240}{{\ttfamily hep-ph/0401240}}].

\bibitem{Hahn-Woernle:2008tsk}
F.~Hahn-Woernle and M.~Plumacher, \emph{{Effects of reheating on
  leptogenesis}},
  \href{https://doi.org/10.1016/j.nuclphysb.2008.07.032}{\emph{Nucl. Phys. B}
  {\bfseries 806} (2009) 68} [\href{https://arxiv.org/abs/0801.3972}{{\ttfamily
  0801.3972}}].

\bibitem{Davidson:2008bu}
S.~Davidson, E.~Nardi and Y.~Nir, \emph{{Leptogenesis}},
  \href{https://doi.org/10.1016/j.physrep.2008.06.002}{\emph{Phys. Rept.}
  {\bfseries 466} (2008) 105}
  [\href{https://arxiv.org/abs/0802.2962}{{\ttfamily 0802.2962}}].

\bibitem{Garcia:2017tuj}
M.A.G.~García, Y.~Mambrini, K.A.~Olive and M.~Peloso, \emph{{Enhancement of
  the Dark Matter Abundance Before Reheating: Applications to Gravitino Dark
  Matter}}, \href{https://doi.org/10.1103/PhysRevD.96.103510}{\emph{Phys. Rev.
  D} {\bfseries 96} (2017) 103510}
  [\href{https://arxiv.org/abs/1709.01549}{{\ttfamily 1709.01549}}].

\bibitem{Bernal:2018qlk}
N.~Bernal, M.~Dutra, Y.~Mambrini, K.~Olive, M.~Peloso and M.~Pierre,
  \emph{{Spin-2 Portal Dark Matter}},
  \href{https://doi.org/10.1103/PhysRevD.97.115020}{\emph{Phys. Rev. D}
  {\bfseries 97} (2018) 115020}
  [\href{https://arxiv.org/abs/1803.01866}{{\ttfamily 1803.01866}}].

\bibitem{Bernal:2018ins}
N.~Bernal, C.~Cosme and T.~Tenkanen, \emph{{Phenomenology of Self-Interacting
  Dark Matter in a Matter-Dominated Universe}},
  \href{https://doi.org/10.1140/epjc/s10052-019-6608-8}{\emph{Eur. Phys. J. C}
  {\bfseries 79} (2019) 99} [\href{https://arxiv.org/abs/1803.08064}{{\ttfamily
  1803.08064}}].

\bibitem{Almeida:2018oid}
J.P.B.~Almeida, N.~Bernal, J.~Rubio and T.~Tenkanen, \emph{{Hidden inflation
  dark matter}},
  \href{https://doi.org/10.1088/1475-7516/2019/03/012}{\emph{JCAP} {\bfseries
  03} (2019) 012} [\href{https://arxiv.org/abs/1811.09640}{{\ttfamily
  1811.09640}}].

\bibitem{Kaneta:2019zgw}
K.~Kaneta, Y.~Mambrini and K.A.~Olive, \emph{{Radiative production of
  nonthermal dark matter}},
  \href{https://doi.org/10.1103/PhysRevD.99.063508}{\emph{Phys. Rev. D}
  {\bfseries 99} (2019) 063508}
  [\href{https://arxiv.org/abs/1901.04449}{{\ttfamily 1901.04449}}].

\bibitem{Bernal:2019mhf}
N.~Bernal, F.~Elahi, C.~Maldonado and J.~Unwin, \emph{{Ultraviolet Freeze-in
  and Non-Standard Cosmologies}},
  \href{https://doi.org/10.1088/1475-7516/2019/11/026}{\emph{JCAP} {\bfseries
  11} (2019) 026} [\href{https://arxiv.org/abs/1909.07992}{{\ttfamily
  1909.07992}}].

\bibitem{Garcia:2020wiy}
M.A.G.~García, K.~Kaneta, Y.~Mambrini and K.A.~Olive, \emph{{Inflaton
  Oscillations and Post-Inflationary Reheating}},
  \href{https://doi.org/10.1088/1475-7516/2021/04/012}{\emph{JCAP} {\bfseries
  04} (2021) 012} [\href{https://arxiv.org/abs/2012.10756}{{\ttfamily
  2012.10756}}].

\bibitem{Mambrini:2021zpp}
Y.~Mambrini and K.A.~Olive, \emph{{Gravitational Production of Dark Matter
  during Reheating}},
  \href{https://doi.org/10.1103/PhysRevD.103.115009}{\emph{Phys. Rev. D}
  {\bfseries 103} (2021) 115009}
  [\href{https://arxiv.org/abs/2102.06214}{{\ttfamily 2102.06214}}].

\bibitem{Bernal:2021kaj}
N.~Bernal and C.S.~Fong, \emph{{Dark matter and leptogenesis from gravitational
  production}},
  \href{https://doi.org/10.1088/1475-7516/2021/06/028}{\emph{JCAP} {\bfseries
  06} (2021) 028} [\href{https://arxiv.org/abs/2103.06896}{{\ttfamily
  2103.06896}}].

\bibitem{Barman:2021ugy}
B.~Barman and N.~Bernal, \emph{{Gravitational SIMPs}},
  \href{https://doi.org/10.1088/1475-7516/2021/06/011}{\emph{JCAP} {\bfseries
  06} (2021) 011} [\href{https://arxiv.org/abs/2104.10699}{{\ttfamily
  2104.10699}}].

\bibitem{Kaneta:2021pyx}
K.~Kaneta, P.~Ko and W.-I.~Park, \emph{{Conformal portal to dark matter}},
  \href{https://doi.org/10.1103/PhysRevD.104.075018}{\emph{Phys. Rev. D}
  {\bfseries 104} (2021) 075018}
  [\href{https://arxiv.org/abs/2106.01923}{{\ttfamily 2106.01923}}].

\bibitem{Clery:2021bwz}
S.~Cl\'ery, Y.~Mambrini, K.A.~Olive and S.~Verner, \emph{{Gravitational portals
  in the early Universe}},
  \href{https://doi.org/10.1103/PhysRevD.105.075005}{\emph{Phys. Rev. D}
  {\bfseries 105} (2022) 075005}
  [\href{https://arxiv.org/abs/2112.15214}{{\ttfamily 2112.15214}}].

\bibitem{Haque:2022kez}
M.R.~Haque and D.~Maity, \emph{{Gravitational reheating}},
  \href{https://doi.org/10.1103/PhysRevD.107.043531}{\emph{Phys. Rev. D}
  {\bfseries 107} (2023) 043531}
  [\href{https://arxiv.org/abs/2201.02348}{{\ttfamily 2201.02348}}].

\bibitem{Barman:2022tzk}
B.~Barman, N.~Bernal, Y.~Xu and {\'O}.~Zapata, \emph{{Ultraviolet freeze-in
  with a time-dependent inflaton decay}},
  \href{https://doi.org/10.1088/1475-7516/2022/07/019}{\emph{JCAP} {\bfseries
  07} (2022) 019} [\href{https://arxiv.org/abs/2202.12906}{{\ttfamily
  2202.12906}}].

\bibitem{Clery:2022wib}
S.~Cl\'ery, Y.~Mambrini, K.A.~Olive, A.~Shkerin and S.~Verner,
  \emph{{Gravitational portals with nonminimal couplings}},
  \href{https://doi.org/10.1103/PhysRevD.105.095042}{\emph{Phys. Rev. D}
  {\bfseries 105} (2022) 095042}
  [\href{https://arxiv.org/abs/2203.02004}{{\ttfamily 2203.02004}}].

\bibitem{Bernal:2022wck}
N.~Bernal and Y.~Xu, \emph{{WIMPs during reheating}},
  \href{https://doi.org/10.1088/1475-7516/2022/12/017}{\emph{JCAP} {\bfseries
  12} (2022) 017} [\href{https://arxiv.org/abs/2209.07546}{{\ttfamily
  2209.07546}}].

\bibitem{Haque:2023yra}
M.R.~Haque, D.~Maity and R.~Mondal, \emph{{WIMPs, FIMPs, and Inflaton
  phenomenology via reheating, CMB and \ensuremath{\Delta}N$_{eff}$}},
  \href{https://doi.org/10.1007/JHEP09(2023)012}{\emph{JHEP} {\bfseries 09}
  (2023) 012} [\href{https://arxiv.org/abs/2301.01641}{{\ttfamily
  2301.01641}}].

\bibitem{Datta:2023pav}
A.~Datta, R.~Roshan and A.~Sil, \emph{{Flavor leptogenesis during the reheating
  era}}, \href{https://doi.org/10.1103/PhysRevD.108.035029}{\emph{Phys. Rev. D}
  {\bfseries 108} (2023) 035029}
  [\href{https://arxiv.org/abs/2301.10791}{{\ttfamily 2301.10791}}].

\bibitem{Silva-Malpartida:2023yks}
J.~Silva-Malpartida, N.~Bernal, J.~Jones-P\'erez and R.A.~Lineros, \emph{{From
  WIMPs to FIMPs with low~reheating~temperatures}},
  \href{https://doi.org/10.1088/1475-7516/2023/09/015}{\emph{JCAP} {\bfseries
  09} (2023) 015} [\href{https://arxiv.org/abs/2306.14943}{{\ttfamily
  2306.14943}}].

\bibitem{Becker:2023tvd}
M.~Becker, E.~Copello, J.~Harz, J.~Lang and Y.~Xu, \emph{{Confronting dark
  matter freeze-in during reheating with constraints from inflation}},
  \href{https://doi.org/10.1088/1475-7516/2024/01/053}{\emph{JCAP} {\bfseries
  01} (2024) 053} [\href{https://arxiv.org/abs/2306.17238}{{\ttfamily
  2306.17238}}].

\bibitem{Banerjee:2024caa}
A.~Banerjee, D.~Chowdhury, A.~Hait and M.S.~Islam, \emph{{Dark matter cooling
  during early matter-domination boosts sub-earth halos}},
  \href{https://doi.org/10.1088/1475-7516/2025/03/030}{\emph{JCAP} {\bfseries
  03} (2025) 030} [\href{https://arxiv.org/abs/2408.08360}{{\ttfamily
  2408.08360}}].

\bibitem{Barman:2024mqo}
B.~Barman, N.~Bernal and Y.~Xu, \emph{{Resonant reheating}},
  \href{https://doi.org/10.1088/1475-7516/2024/08/014}{\emph{JCAP} {\bfseries
  08} (2024) 014} [\href{https://arxiv.org/abs/2404.16090}{{\ttfamily
  2404.16090}}].

\bibitem{Barman:2024ujh}
B.~Barman, A.~Basu, D.~Borah, A.~Chakraborty and R.~Roshan, \emph{{Testing
  leptogenesis and dark matter production during reheating with primordial
  gravitational waves}},
  \href{https://doi.org/10.1103/PhysRevD.111.055016}{\emph{Phys. Rev. D}
  {\bfseries 111} (2025) 055016}
  [\href{https://arxiv.org/abs/2410.19048}{{\ttfamily 2410.19048}}].

\bibitem{Bernal:2024ndy}
N.~Bernal, C.S.~Fong and {\'O}.~Zapata, \emph{{Probing low-reheating scenarios
  with minimal freeze-in dark matter}},
  \href{https://doi.org/10.1007/JHEP02(2025)161}{\emph{JHEP} {\bfseries 02}
  (2025) 161} [\href{https://arxiv.org/abs/2412.04550}{{\ttfamily
  2412.04550}}].

\bibitem{Belanger:2024yoj}
G.~B\'elanger, N.~Bernal and A.~Pukhov, \emph{{Z'-mediated dark matter with
  low-temperature reheating}},
  \href{https://doi.org/10.1007/JHEP03(2025)079}{\emph{JHEP} {\bfseries 03}
  (2025) 079} [\href{https://arxiv.org/abs/2412.12303}{{\ttfamily
  2412.12303}}].

\bibitem{Berera:1995ie}
A.~Berera, \emph{{Warm inflation}},
  \href{https://doi.org/10.1103/PhysRevLett.75.3218}{\emph{Phys. Rev. Lett.}
  {\bfseries 75} (1995) 3218}
  [\href{https://arxiv.org/abs/astro-ph/9509049}{{\ttfamily
  astro-ph/9509049}}].

\bibitem{Berera:1996nv}
A.~Berera, \emph{{Thermal properties of an inflationary universe}},
  \href{https://doi.org/10.1103/PhysRevD.54.2519}{\emph{Phys. Rev. D}
  {\bfseries 54} (1996) 2519}
  [\href{https://arxiv.org/abs/hep-th/9601134}{{\ttfamily hep-th/9601134}}].

\bibitem{Berera:2008ar}
A.~Berera, I.G.~Moss and R.O.~Ramos, \emph{{Warm Inflation and its
  Microphysical Basis}},
  \href{https://doi.org/10.1088/0034-4885/72/2/026901}{\emph{Rept. Prog. Phys.}
  {\bfseries 72} (2009) 026901}
  [\href{https://arxiv.org/abs/0808.1855}{{\ttfamily 0808.1855}}].

\bibitem{Bastero-Gil:2009sdq}
M.~Bastero-Gil and A.~Berera, \emph{{Warm inflation model building}},
  \href{https://doi.org/10.1142/S0217751X09044206}{\emph{Int. J. Mod. Phys. A}
  {\bfseries 24} (2009) 2207}
  [\href{https://arxiv.org/abs/0902.0521}{{\ttfamily 0902.0521}}].

\bibitem{Bastero-Gil:2010dgy}
M.~Bastero-Gil, A.~Berera and R.O.~Ramos, \emph{{Dissipation coefficients from
  scalar and fermion quantum field interactions}},
  \href{https://doi.org/10.1088/1475-7516/2011/09/033}{\emph{JCAP} {\bfseries
  09} (2011) 033} [\href{https://arxiv.org/abs/1008.1929}{{\ttfamily
  1008.1929}}].

\bibitem{Bastero-Gil:2016qru}
M.~Bastero-Gil, A.~Berera, R.O.~Ramos and J.G.~Rosa, \emph{{Warm Little
  Inflaton}}, \href{https://doi.org/10.1103/PhysRevLett.117.151301}{\emph{Phys.
  Rev. Lett.} {\bfseries 117} (2016) 151301}
  [\href{https://arxiv.org/abs/1604.08838}{{\ttfamily 1604.08838}}].

\bibitem{Planck:2018jri}
{\scshape Planck} collaboration, \emph{{Planck 2018 results. X. Constraints on
  inflation}}, \href{https://doi.org/10.1051/0004-6361/201833887}{\emph{Astron.
  Astrophys.} {\bfseries 641} (2020) A10}
  [\href{https://arxiv.org/abs/1807.06211}{{\ttfamily 1807.06211}}].

\bibitem{Starobinsky:1980te}
A.A.~Starobinsky, \emph{{A New Type of Isotropic Cosmological Models Without
  Singularity}},
  \href{https://doi.org/10.1016/0370-2693(80)90670-X}{\emph{Phys. Lett. B}
  {\bfseries 91} (1980) 99}.

\bibitem{Starobinsky:1981vz}
A.A.~Starobinsky, \emph{{Nonsingular Model of the Universe with the Quantum
  Gravitational de Sitter Stage and its Observational Consequences}},  in
  \emph{{Second Seminar on Quantum Gravity}}, 1981.

\bibitem{Starobinsky:1983zz}
A.A.~Starobinsky, \emph{{The Perturbation Spectrum Evolving from a Nonsingular
  Initially De-Sitter Cosmology and the Microwave Background Anisotropy}},
  {\emph{Sov. Astron. Lett.} {\bfseries 9} (1983) 302}.

\bibitem{Kofman:1985aw}
L.A.~Kofman, A.D.~Linde and A.A.~Starobinsky, \emph{{Inflationary Universe
  Generated by the Combined Action of a Scalar Field and Gravitational Vacuum
  Polarization}},
  \href{https://doi.org/10.1016/0370-2693(85)90381-8}{\emph{Phys. Lett. B}
  {\bfseries 157} (1985) 361}.

\bibitem{Rubio:2018ogq}
J.~Rubio, \emph{{Higgs inflation}},
  \href{https://doi.org/10.3389/fspas.2018.00050}{\emph{Front. Astron. Space
  Sci.} {\bfseries 5} (2019) 50}
  [\href{https://arxiv.org/abs/1807.02376}{{\ttfamily 1807.02376}}].

\bibitem{Garcia-Bellido:2011kqb}
J.~García-Bellido, J.~Rubio, M.~Shaposhnikov and D.~Zenhausern,
  \emph{{Higgs-Dilaton Cosmology: From the Early to the Late Universe}},
  \href{https://doi.org/10.1103/PhysRevD.84.123504}{\emph{Phys. Rev. D}
  {\bfseries 84} (2011) 123504}
  [\href{https://arxiv.org/abs/1107.2163}{{\ttfamily 1107.2163}}].

\bibitem{Karananas:2016kyt}
G.K.~Karananas and J.~Rubio, \emph{{On the geometrical interpretation of
  scale-invariant models of inflation}},
  \href{https://doi.org/10.1016/j.physletb.2016.08.037}{\emph{Phys. Lett. B}
  {\bfseries 761} (2016) 223}
  [\href{https://arxiv.org/abs/1606.08848}{{\ttfamily 1606.08848}}].

\bibitem{Casas:2017wjh}
S.~Casas, M.~Pauly and J.~Rubio, \emph{{Higgs-dilaton cosmology: An
  inflation\textendash{}dark-energy connection and forecasts for future galaxy
  surveys}}, \href{https://doi.org/10.1103/PhysRevD.97.043520}{\emph{Phys. Rev.
  D} {\bfseries 97} (2018) 043520}
  [\href{https://arxiv.org/abs/1712.04956}{{\ttfamily 1712.04956}}].

\bibitem{Casas:2018fum}
S.~Casas, G.K.~Karananas, M.~Pauly and J.~Rubio, \emph{{Scale-invariant
  alternatives to general relativity. III. The inflation-dark energy
  connection}}, \href{https://doi.org/10.1103/PhysRevD.99.063512}{\emph{Phys.
  Rev. D} {\bfseries 99} (2019) 063512}
  [\href{https://arxiv.org/abs/1811.05984}{{\ttfamily 1811.05984}}].

\bibitem{Piani:2022gon}
M.~Piani and J.~Rubio, \emph{{Higgs-Dilaton inflation in Einstein-Cartan
  gravity}}, \href{https://doi.org/10.1088/1475-7516/2022/05/009}{\emph{JCAP}
  {\bfseries 05} (2022) 009}
  [\href{https://arxiv.org/abs/2202.04665}{{\ttfamily 2202.04665}}].

\bibitem{Kallosh:2013hoa}
R.~Kallosh and A.~Linde, \emph{{Universality Class in Conformal Inflation}},
  \href{https://doi.org/10.1088/1475-7516/2013/07/002}{\emph{JCAP} {\bfseries
  07} (2013) 002} [\href{https://arxiv.org/abs/1306.5220}{{\ttfamily
  1306.5220}}].

\bibitem{Artymowski:2016pjz}
M.~Artymowski and J.~Rubio, \emph{{Endlessly flat scalar potentials and
  $\alpha$-attractors}},
  \href{https://doi.org/10.1016/j.physletb.2016.08.024}{\emph{Phys. Lett. B}
  {\bfseries 761} (2016) 111}
  [\href{https://arxiv.org/abs/1607.00398}{{\ttfamily 1607.00398}}].

\bibitem{Linde:1983gd}
A.D.~Linde, \emph{{Chaotic Inflation}},
  \href{https://doi.org/10.1016/0370-2693(83)90837-7}{\emph{Phys. Lett. B}
  {\bfseries 129} (1983) 177}.

\bibitem{Ellis:2015kqa}
J.~Ellis, M.A.G.~Garc\'ia, D.V.~Nanopoulos and K.A.~Olive,
  \emph{{Phenomenological Aspects of No-Scale Inflation Models}},
  \href{https://doi.org/10.1088/1475-7516/2015/10/003}{\emph{JCAP} {\bfseries
  10} (2015) 003} [\href{https://arxiv.org/abs/1503.08867}{{\ttfamily
  1503.08867}}].

\bibitem{Ellis:2015pla}
J.~Ellis, M.A.G.~Garc\'ia, D.V.~Nanopoulos and K.A.~Olive, \emph{{Calculations
  of Inflaton Decays and Reheating: with Applications to No-Scale Inflation
  Models}}, \href{https://doi.org/10.1088/1475-7516/2015/07/050}{\emph{JCAP}
  {\bfseries 07} (2015) 050}
  [\href{https://arxiv.org/abs/1505.06986}{{\ttfamily 1505.06986}}].

\bibitem{Choi:1994ax}
S.Y.~Choi, J.S.~Shim and H.S.~Song, \emph{{Factorization and polarization in
  linearized gravity}},
  \href{https://doi.org/10.1103/PhysRevD.51.2751}{\emph{Phys. Rev. D}
  {\bfseries 51} (1995) 2751}
  [\href{https://arxiv.org/abs/hep-th/9411092}{{\ttfamily hep-th/9411092}}].

\bibitem{Garny:2015sjg}
M.~Garny, M.~Sandora and M.S.~Sloth, \emph{{Planckian Interacting Massive
  Particles as Dark Matter}},
  \href{https://doi.org/10.1103/PhysRevLett.116.101302}{\emph{Phys. Rev. Lett.}
  {\bfseries 116} (2016) 101302}
  [\href{https://arxiv.org/abs/1511.03278}{{\ttfamily 1511.03278}}].

\bibitem{Tang:2017hvq}
Y.~Tang and Y.-L.~Wu, \emph{{On Thermal Gravitational Contribution to Particle
  Production and Dark Matter}},
  \href{https://doi.org/10.1016/j.physletb.2017.10.034}{\emph{Phys. Lett. B}
  {\bfseries 774} (2017) 676}
  [\href{https://arxiv.org/abs/1708.05138}{{\ttfamily 1708.05138}}].

\bibitem{Garny:2017kha}
M.~Garny, A.~Palessandro, M.~Sandora and M.S.~Sloth, \emph{{Theory and
  Phenomenology of Planckian Interacting Massive Particles as Dark Matter}},
  \href{https://doi.org/10.1088/1475-7516/2018/02/027}{\emph{JCAP} {\bfseries
  02} (2018) 027} [\href{https://arxiv.org/abs/1709.09688}{{\ttfamily
  1709.09688}}].

\bibitem{Ahmed:2020fhc}
A.~Ahmed, B.~Grzadkowski and A.~Socha, \emph{{Gravitational production of
  vector dark matter}},
  \href{https://doi.org/10.1007/JHEP08(2020)059}{\emph{JHEP} {\bfseries 08}
  (2020) 059} [\href{https://arxiv.org/abs/2005.01766}{{\ttfamily
  2005.01766}}].

\bibitem{Co:2022bgh}
R.T.~Co, Y.~Mambrini and K.A.~Olive, \emph{{Inflationary gravitational
  leptogenesis}},
  \href{https://doi.org/10.1103/PhysRevD.106.075006}{\emph{Phys. Rev. D}
  {\bfseries 106} (2022) 075006}
  [\href{https://arxiv.org/abs/2205.01689}{{\ttfamily 2205.01689}}].

\bibitem{Barman:2022qgt}
B.~Barman, S.~Cl\'ery, R.T.~Co, Y.~Mambrini and K.A.~Olive, \emph{{Gravity as a
  portal to reheating, leptogenesis and dark matter}},
  \href{https://doi.org/10.1007/JHEP12(2022)072}{\emph{JHEP} {\bfseries 12}
  (2022) 072} [\href{https://arxiv.org/abs/2210.05716}{{\ttfamily
  2210.05716}}].

\bibitem{Barman:2023opy}
B.~Barman, N.~Bernal and J.~Rubio, \emph{{Rescuing gravitational-reheating in
  chaotic inflation}},
  \href{https://doi.org/10.1088/1475-7516/2024/05/072}{\emph{JCAP} {\bfseries
  05} (2024) 072} [\href{https://arxiv.org/abs/2310.06039}{{\ttfamily
  2310.06039}}].

\bibitem{Garcia:2020eof}
M.A.G.~García, K.~Kaneta, Y.~Mambrini and K.A.~Olive, \emph{{Reheating and
  Post-inflationary Production of Dark Matter}},
  \href{https://doi.org/10.1103/PhysRevD.101.123507}{\emph{Phys. Rev. D}
  {\bfseries 101} (2020) 123507}
  [\href{https://arxiv.org/abs/2004.08404}{{\ttfamily 2004.08404}}].

\bibitem{Ahmed:2021fvt}
A.~Ahmed, B.~Grzadkowski and A.~Socha, \emph{{Implications of time-dependent
  inflaton decay on reheating and dark matter production}},
  \href{https://doi.org/10.1016/j.physletb.2022.137201}{\emph{Phys. Lett. B}
  {\bfseries 831} (2022) 137201}
  [\href{https://arxiv.org/abs/2111.06065}{{\ttfamily 2111.06065}}].

\bibitem{Ichikawa:2008ne}
K.~Ichikawa, T.~Suyama, T.~Takahashi and M.~Yamaguchi, \emph{{Primordial
  Curvature Fluctuation and Its Non-Gaussianity in Models with Modulated
  Reheating}}, \href{https://doi.org/10.1103/PhysRevD.78.063545}{\emph{Phys.
  Rev. D} {\bfseries 78} (2008) 063545}
  [\href{https://arxiv.org/abs/0807.3988}{{\ttfamily 0807.3988}}].

\bibitem{Kainulainen:2016vzv}
K.~Kainulainen, S.~Nurmi, T.~Tenkanen, K.~Tuominen and V.~Vaskonen,
  \emph{{Isocurvature Constraints on Portal Couplings}},
  \href{https://doi.org/10.1088/1475-7516/2016/06/022}{\emph{JCAP} {\bfseries
  06} (2016) 022} [\href{https://arxiv.org/abs/1601.07733}{{\ttfamily
  1601.07733}}].

\bibitem{Ahmed:2022tfm}
A.~Ahmed, B.~Grzadkowski and A.~Socha, \emph{{Higgs boson induced reheating and
  ultraviolet frozen-in dark matter}},
  \href{https://doi.org/10.1007/JHEP02(2023)196}{\emph{JHEP} {\bfseries 02}
  (2023) 196} [\href{https://arxiv.org/abs/2207.11218}{{\ttfamily
  2207.11218}}].

\bibitem{Shtanov:1994ce}
Y.~Shtanov, J.H.~Traschen and R.H.~Brandenberger, \emph{{Universe reheating
  after inflation}},
  \href{https://doi.org/10.1103/PhysRevD.51.5438}{\emph{Phys. Rev. D}
  {\bfseries 51} (1995) 5438}
  [\href{https://arxiv.org/abs/hep-ph/9407247}{{\ttfamily hep-ph/9407247}}].

\bibitem{Turner:1983he}
M.S.~Turner, \emph{{Coherent Scalar Field Oscillations in an Expanding
  Universe}}, \href{https://doi.org/10.1103/PhysRevD.28.1243}{\emph{Phys. Rev.
  D} {\bfseries 28} (1983) 1243}.

\bibitem{Johnson:2008se}
M.C.~Johnson and M.~Kamionkowski, \emph{{Dynamical and Gravitational
  Instability of Oscillating-Field Dark Energy and Dark Matter}},
  \href{https://doi.org/10.1103/PhysRevD.78.063010}{\emph{Phys. Rev. D}
  {\bfseries 78} (2008) 063010}
  [\href{https://arxiv.org/abs/0805.1748}{{\ttfamily 0805.1748}}].

\bibitem{Davidson:2000er}
S.~Davidson and S.~Sarkar, \emph{{Thermalization after inflation}},
  \href{https://doi.org/10.1088/1126-6708/2000/11/012}{\emph{JHEP} {\bfseries
  11} (2000) 012} [\href{https://arxiv.org/abs/hep-ph/0009078}{{\ttfamily
  hep-ph/0009078}}].

\bibitem{Allahverdi:2002pu}
R.~Allahverdi and M.~Drees, \emph{{Thermalization after inflation and
  production of massive stable particles}},
  \href{https://doi.org/10.1103/PhysRevD.66.063513}{\emph{Phys. Rev. D}
  {\bfseries 66} (2002) 063513}
  [\href{https://arxiv.org/abs/hep-ph/0205246}{{\ttfamily hep-ph/0205246}}].

\bibitem{Kurkela:2011ti}
A.~Kurkela and G.D.~Moore, \emph{{Thermalization in Weakly Coupled Nonabelian
  Plasmas}}, \href{https://doi.org/10.1007/JHEP12(2011)044}{\emph{JHEP}
  {\bfseries 12} (2011) 044} [\href{https://arxiv.org/abs/1107.5050}{{\ttfamily
  1107.5050}}].

\bibitem{Harigaya:2013vwa}
K.~Harigaya and K.~Mukaida, \emph{{Thermalization after/during Reheating}},
  \href{https://doi.org/10.1007/JHEP05(2014)006}{\emph{JHEP} {\bfseries 05}
  (2014) 006} [\href{https://arxiv.org/abs/1312.3097}{{\ttfamily 1312.3097}}].

\bibitem{Giudice:2000ex}
G.F.~Giudice, E.W.~Kolb and A.~Riotto, \emph{{Largest temperature of the
  radiation era and its cosmological implications}},
  \href{https://doi.org/10.1103/PhysRevD.64.023508}{\emph{Phys. Rev. D}
  {\bfseries 64} (2001) 023508}
  [\href{https://arxiv.org/abs/hep-ph/0005123}{{\ttfamily hep-ph/0005123}}].

\bibitem{Sarkar:1995dd}
S.~Sarkar, \emph{{Big bang nucleosynthesis and physics beyond the standard
  model}}, \href{https://doi.org/10.1088/0034-4885/59/12/001}{\emph{Rept. Prog.
  Phys.} {\bfseries 59} (1996) 1493}
  [\href{https://arxiv.org/abs/hep-ph/9602260}{{\ttfamily hep-ph/9602260}}].

\bibitem{Kawasaki:2000en}
M.~Kawasaki, K.~Kohri and N.~Sugiyama, \emph{{MeV scale reheating temperature
  and thermalization of neutrino background}},
  \href{https://doi.org/10.1103/PhysRevD.62.023506}{\emph{Phys. Rev. D}
  {\bfseries 62} (2000) 023506}
  [\href{https://arxiv.org/abs/astro-ph/0002127}{{\ttfamily
  astro-ph/0002127}}].

\bibitem{Hannestad:2004px}
S.~Hannestad, \emph{{What is the lowest possible reheating temperature?}},
  \href{https://doi.org/10.1103/PhysRevD.70.043506}{\emph{Phys. Rev. D}
  {\bfseries 70} (2004) 043506}
  [\href{https://arxiv.org/abs/astro-ph/0403291}{{\ttfamily
  astro-ph/0403291}}].

\bibitem{DeBernardis:2008zz}
F.~De~Bernardis, L.~Pagano and A.~Melchiorri, \emph{{New constraints on the
  reheating temperature of the universe after WMAP-5}},
  \href{https://doi.org/10.1016/j.astropartphys.2008.09.005}{\emph{Astropart.
  Phys.} {\bfseries 30} (2008) 192}.

\bibitem{deSalas:2015glj}
P.F.~de~Salas, M.~Lattanzi, G.~Mangano, G.~Miele, S.~Pastor and O.~Pisanti,
  \emph{{Bounds on very low reheating scenarios after Planck}},
  \href{https://doi.org/10.1103/PhysRevD.92.123534}{\emph{Phys. Rev. D}
  {\bfseries 92} (2015) 123534}
  [\href{https://arxiv.org/abs/1511.00672}{{\ttfamily 1511.00672}}].

\bibitem{Hasegawa:2019jsa}
T.~Hasegawa, N.~Hiroshima, K.~Kohri, R.S.L.~Hansen, T.~Tram and S.~Hannestad,
  \emph{{MeV-scale reheating temperature and thermalization of oscillating
  neutrinos by radiative and hadronic decays of massive particles}},
  \href{https://doi.org/10.1088/1475-7516/2019/12/012}{\emph{JCAP} {\bfseries
  12} (2019) 012} [\href{https://arxiv.org/abs/1908.10189}{{\ttfamily
  1908.10189}}].

\bibitem{Barbieri:2025moq}
N.~Barbieri, T.~Brinckmann, S.~Gariazzo, M.~Lattanzi, S.~Pastor and O.~Pisanti,
  \emph{{Current constraints on cosmological scenarios with very low reheating
  temperatures}},  \href{https://arxiv.org/abs/2501.01369}{{\ttfamily
  2501.01369}}.

\bibitem{Co:2020xaf}
R.T.~Co, E.~Gonz\'alez and K.~Harigaya, \emph{{Increasing Temperature toward
  the Completion of Reheating}},
  \href{https://doi.org/10.1088/1475-7516/2020/11/038}{\emph{JCAP} {\bfseries
  11} (2020) 038} [\href{https://arxiv.org/abs/2007.04328}{{\ttfamily
  2007.04328}}].

\bibitem{Chowdhury:2023jft}
D.~Chowdhury and A.~Hait, \emph{{Thermalization in the presence of a
  time-dependent dissipation and its impact on dark matter production}},
  \href{https://doi.org/10.1007/JHEP09(2023)085}{\emph{JHEP} {\bfseries 09}
  (2023) 085} [\href{https://arxiv.org/abs/2302.06654}{{\ttfamily
  2302.06654}}].

\bibitem{Cosme:2024ndc}
C.~Cosme, F.~Costa and O.~Lebedev, \emph{{Temperature evolution in the Early
  Universe and freeze-in at stronger coupling}},
  \href{https://doi.org/10.1088/1475-7516/2024/06/031}{\emph{JCAP} {\bfseries
  06} (2024) 031} [\href{https://arxiv.org/abs/2402.04743}{{\ttfamily
  2402.04743}}].

\bibitem{Spokoiny:1993kt}
B.~Spokoiny, \emph{{Deflationary universe scenario}},
  \href{https://doi.org/10.1016/0370-2693(93)90155-B}{\emph{Phys. Lett. B}
  {\bfseries 315} (1993) 40}
  [\href{https://arxiv.org/abs/gr-qc/9306008}{{\ttfamily gr-qc/9306008}}].

\bibitem{Ferreira:1997hj}
P.G.~Ferreira and M.~Joyce, \emph{{Cosmology with a primordial scaling field}},
  \href{https://doi.org/10.1103/PhysRevD.58.023503}{\emph{Phys. Rev. D}
  {\bfseries 58} (1998) 023503}
  [\href{https://arxiv.org/abs/astro-ph/9711102}{{\ttfamily
  astro-ph/9711102}}].

\bibitem{Nakayama:2018ptw}
K.~Nakayama and Y.~Tang, \emph{{Stochastic Gravitational Waves from Particle
  Origin}}, \href{https://doi.org/10.1016/j.physletb.2018.11.023}{\emph{Phys.
  Lett. B} {\bfseries 788} (2019) 341}
  [\href{https://arxiv.org/abs/1810.04975}{{\ttfamily 1810.04975}}].

\bibitem{Huang:2019lgd}
D.~Huang and L.~Yin, \emph{{Stochastic Gravitational Waves from Inflaton
  Decays}}, \href{https://doi.org/10.1103/PhysRevD.100.043538}{\emph{Phys. Rev.
  D} {\bfseries 100} (2019) 043538}
  [\href{https://arxiv.org/abs/1905.08510}{{\ttfamily 1905.08510}}].

\bibitem{Barman:2023ymn}
B.~Barman, N.~Bernal, Y.~Xu and O.~Zapata, \emph{{Gravitational wave from
  graviton Bremsstrahlung during reheating}},
  \href{https://doi.org/10.1088/1475-7516/2023/05/019}{\emph{JCAP} {\bfseries
  05} (2023) 019} [\href{https://arxiv.org/abs/2301.11345}{{\ttfamily
  2301.11345}}].

\bibitem{Barman:2023rpg}
B.~Barman, N.~Bernal, Y.~Xu and {\'O}.~Zapata, \emph{{Bremsstrahlung-induced
  gravitational waves in monomial potentials during reheating}},
  \href{https://doi.org/10.1103/PhysRevD.108.083524}{\emph{Phys. Rev. D}
  {\bfseries 108} (2023) 083524}
  [\href{https://arxiv.org/abs/2305.16388}{{\ttfamily 2305.16388}}].

\bibitem{Kanemura:2023pnv}
S.~Kanemura and K.~Kaneta, \emph{{Gravitational waves from particle decays
  during reheating}},
  \href{https://doi.org/10.1016/j.physletb.2024.138807}{\emph{Phys. Lett. B}
  {\bfseries 855} (2024) 138807}
  [\href{https://arxiv.org/abs/2310.12023}{{\ttfamily 2310.12023}}].

\bibitem{Bernal:2023wus}
N.~Bernal, S.~Cl\'ery, Y.~Mambrini and Y.~Xu, \emph{{Probing reheating with
  graviton bremsstrahlung}},
  \href{https://doi.org/10.1088/1475-7516/2024/01/065}{\emph{JCAP} {\bfseries
  01} (2024) 065} [\href{https://arxiv.org/abs/2311.12694}{{\ttfamily
  2311.12694}}].

\bibitem{Tokareva:2023mrt}
A.~Tokareva, \emph{{Gravitational waves from inflaton decay and
  bremsstrahlung}},
  \href{https://doi.org/10.1016/j.physletb.2024.138695}{\emph{Phys. Lett. B}
  {\bfseries 853} (2024) 138695}
  [\href{https://arxiv.org/abs/2312.16691}{{\ttfamily 2312.16691}}].

\bibitem{Hu:2024awd}
W.~Hu, K.~Nakayama, V.~Takhistov and Y.~Tang, \emph{{Gravitational wave probe
  of Planck-scale physics after inflation}},
  \href{https://doi.org/10.1016/j.physletb.2024.138958}{\emph{Phys. Lett. B}
  {\bfseries 856} (2024) 138958}
  [\href{https://arxiv.org/abs/2403.13882}{{\ttfamily 2403.13882}}].

\bibitem{Choi:2024acs}
K.-Y.~Choi, E.~Lkhagvadorj and S.~Mahapatra, \emph{{Gravitational wave sourced
  by decay of massive particle from primordial black hole evaporation}},
  \href{https://doi.org/10.1088/1475-7516/2024/07/064}{\emph{JCAP} {\bfseries
  07} (2024) 064} [\href{https://arxiv.org/abs/2403.15269}{{\ttfamily
  2403.15269}}].

\bibitem{Barman:2024htg}
B.~Barman, N.~Bernal, S.~Cl\'ery, Y.~Mambrini, Y.~Xu and {\'O}.~Zapata,
  \emph{{Probing Reheating with Gravitational Waves from Graviton
  Bremsstrahlung}},  in \emph{{58$^{th}$ Rencontres de Moriond on Electroweak
  Interactions and Unified Theories}}, 5, 2024
  [\href{https://arxiv.org/abs/2405.09620}{{\ttfamily 2405.09620}}].

\bibitem{Inui:2024wgj}
R.~Inui, Y.~Mikura and S.~Yokoyama, \emph{{Gravitational waves from graviton
  bremsstrahlung with kination phase}},
  \href{https://doi.org/10.1103/PhysRevD.111.043511}{\emph{Phys. Rev. D}
  {\bfseries 111} (2025) 043511}
  [\href{https://arxiv.org/abs/2408.10786}{{\ttfamily 2408.10786}}].

\bibitem{Jiang:2024akb}
Y.~Jiang and T.~Suyama, \emph{{Spectrum of high-frequency gravitational waves
  from graviton bremsstrahlung by the decay of inflaton: case with polynomial
  potential}}, \href{https://doi.org/10.1088/1475-7516/2025/02/041}{\emph{JCAP}
  {\bfseries 02} (2025) 041}
  [\href{https://arxiv.org/abs/2410.11175}{{\ttfamily 2410.11175}}].

\bibitem{Ema:2015dka}
Y.~Ema, R.~Jinno, K.~Mukaida and K.~Nakayama, \emph{{Gravitational Effects on
  Inflaton Decay}},
  \href{https://doi.org/10.1088/1475-7516/2015/05/038}{\emph{JCAP} {\bfseries
  05} (2015) 038} [\href{https://arxiv.org/abs/1502.02475}{{\ttfamily
  1502.02475}}].

\bibitem{Ema:2016hlw}
Y.~Ema, R.~Jinno, K.~Mukaida and K.~Nakayama, \emph{{Gravitational particle
  production in oscillating backgrounds and its cosmological implications}},
  \href{https://doi.org/10.1103/PhysRevD.94.063517}{\emph{Phys. Rev. D}
  {\bfseries 94} (2016) 063517}
  [\href{https://arxiv.org/abs/1604.08898}{{\ttfamily 1604.08898}}].

\bibitem{Ema:2020ggo}
Y.~Ema, R.~Jinno and K.~Nakayama, \emph{{High-frequency Graviton from Inflaton
  Oscillation}},
  \href{https://doi.org/10.1088/1475-7516/2020/09/015}{\emph{JCAP} {\bfseries
  09} (2020) 015} [\href{https://arxiv.org/abs/2006.09972}{{\ttfamily
  2006.09972}}].

\bibitem{Choi:2024ilx}
G.~Choi, W.~Ke and K.A.~Olive, \emph{{Minimal production of prompt
  gravitational waves during reheating}},
  \href{https://doi.org/10.1103/PhysRevD.109.083516}{\emph{Phys. Rev. D}
  {\bfseries 109} (2024) 083516}
  [\href{https://arxiv.org/abs/2402.04310}{{\ttfamily 2402.04310}}].

\bibitem{Xu:2024fjl}
Y.~Xu, \emph{{Ultra-high frequency gravitational waves from scattering,
  Bremsstrahlung and decay during reheating}},
  \href{https://doi.org/10.1007/JHEP10(2024)174}{\emph{JHEP} {\bfseries 10}
  (2024) 174} [\href{https://arxiv.org/abs/2407.03256}{{\ttfamily
  2407.03256}}].

\bibitem{Bernal:2025lxp}
N.~Bernal, Q.-f.~Wu, X.-J.~Xu and Y.~Xu, \emph{{Pre-thermalized Gravitational
  Waves}},  \href{https://arxiv.org/abs/2503.10756}{{\ttfamily 2503.10756}}.

\bibitem{Bernal:2024jim}
N.~Bernal and Y.~Xu, \emph{{Thermal gravitational waves during reheating}},
  \href{https://doi.org/10.1007/JHEP01(2025)137}{\emph{JHEP} {\bfseries 01}
  (2025) 137} [\href{https://arxiv.org/abs/2410.21385}{{\ttfamily
  2410.21385}}].

\bibitem{Ghiglieri:2015nfa}
J.~Ghiglieri and M.~Laine, \emph{{Gravitational wave background from Standard
  Model physics: Qualitative features}},
  \href{https://doi.org/10.1088/1475-7516/2015/07/022}{\emph{JCAP} {\bfseries
  07} (2015) 022} [\href{https://arxiv.org/abs/1504.02569}{{\ttfamily
  1504.02569}}].

\bibitem{Ghiglieri:2020mhm}
J.~Ghiglieri, G.~Jackson, M.~Laine and Y.~Zhu, \emph{{Gravitational wave
  background from Standard Model physics: Complete leading order}},
  \href{https://doi.org/10.1007/JHEP07(2020)092}{\emph{JHEP} {\bfseries 07}
  (2020) 092} [\href{https://arxiv.org/abs/2004.11392}{{\ttfamily
  2004.11392}}].

\bibitem{Ringwald:2020ist}
A.~Ringwald, J.~Sch\"utte-Engel and C.~Tamarit, \emph{{Gravitational Waves as a
  Big Bang Thermometer}},
  \href{https://doi.org/10.1088/1475-7516/2021/03/054}{\emph{JCAP} {\bfseries
  03} (2021) 054} [\href{https://arxiv.org/abs/2011.04731}{{\ttfamily
  2011.04731}}].

\bibitem{Ringwald:2022xif}
A.~Ringwald and C.~Tamarit, \emph{{Revealing the cosmic history with
  gravitational waves}},
  \href{https://doi.org/10.1103/PhysRevD.106.063027}{\emph{Phys. Rev. D}
  {\bfseries 106} (2022) 063027}
  [\href{https://arxiv.org/abs/2203.00621}{{\ttfamily 2203.00621}}].

\bibitem{Klose:2022rxh}
P.~Klose, M.~Laine and S.~Procacci, \emph{{Gravitational wave background from
  vacuum and thermal fluctuations during axion-like inflation}},
  \href{https://doi.org/10.1088/1475-7516/2022/12/020}{\emph{JCAP} {\bfseries
  12} (2022) 020} [\href{https://arxiv.org/abs/2210.11710}{{\ttfamily
  2210.11710}}].

\bibitem{Ghiglieri:2022rfp}
J.~Ghiglieri, J.~Sch\"utte-Engel and E.~Speranza, \emph{{Freezing-In
  Gravitational Waves}},  \href{https://arxiv.org/abs/2211.16513}{{\ttfamily
  2211.16513}}.

\bibitem{Drewes:2023oxg}
M.~Drewes, Y.~Georis, J.~Klaric and P.~Klose, \emph{{Upper bound on thermal
  gravitational wave backgrounds from hidden sectors}},
  \href{https://doi.org/10.1088/1475-7516/2024/06/073}{\emph{JCAP} {\bfseries
  06} (2024) 073} [\href{https://arxiv.org/abs/2312.13855}{{\ttfamily
  2312.13855}}].

\bibitem{Ghiglieri:2024ghm}
J.~Ghiglieri, M.~Laine, J.~Sch\"utte-Engel and E.~Speranza,
  \emph{{Double-graviton production from Standard Model plasma}},
  \href{https://doi.org/10.1088/1475-7516/2024/04/062}{\emph{JCAP} {\bfseries
  04} (2024) 062} [\href{https://arxiv.org/abs/2401.08766}{{\ttfamily
  2401.08766}}].

\bibitem{Kofman:1994rk}
L.~Kofman, A.D.~Linde and A.A.~Starobinsky, \emph{{Reheating after inflation}},
  \href{https://doi.org/10.1103/PhysRevLett.73.3195}{\emph{Phys. Rev. Lett.}
  {\bfseries 73} (1994) 3195}
  [\href{https://arxiv.org/abs/hep-th/9405187}{{\ttfamily hep-th/9405187}}].

\bibitem{Kofman:1997yn}
L.~Kofman, A.D.~Linde and A.A.~Starobinsky, \emph{{Towards the theory of
  reheating after inflation}},
  \href{https://doi.org/10.1103/PhysRevD.56.3258}{\emph{Phys. Rev. D}
  {\bfseries 56} (1997) 3258}
  [\href{https://arxiv.org/abs/hep-ph/9704452}{{\ttfamily hep-ph/9704452}}].

\bibitem{Traschen:1990sw}
J.H.~Traschen and R.H.~Brandenberger, \emph{{Particle Production During
  Out-of-equilibrium Phase Transitions}},
  \href{https://doi.org/10.1103/PhysRevD.42.2491}{\emph{Phys. Rev. D}
  {\bfseries 42} (1990) 2491}.

\bibitem{Dolgov:1989us}
A.D.~Dolgov and D.P.~Kirilova, \emph{{On Particle Creation by a Time Dependent
  Scalar Field}}, {\emph{Sov. J. Nucl. Phys.} {\bfseries 51} (1990) 172}.

\bibitem{Kaiser:1995fb}
D.I.~Kaiser, \emph{{Post inflation reheating in an expanding universe}},
  \href{https://doi.org/10.1103/PhysRevD.53.1776}{\emph{Phys. Rev. D}
  {\bfseries 53} (1996) 1776}
  [\href{https://arxiv.org/abs/astro-ph/9507108}{{\ttfamily
  astro-ph/9507108}}].

\bibitem{Kubo:1966fyg}
R.~Kubo, \emph{{The fluctuation-dissipation theorem}},
  \href{https://doi.org/10.1088/0034-4885/29/1/306}{\emph{Rept. Prog. Phys.}
  {\bfseries 29} (1966) 255}.

\bibitem{Gleiser:1993ea}
M.~Gleiser and R.O.~Ramos, \emph{{Microphysical approach to nonequilibrium
  dynamics of quantum fields}},
  \href{https://doi.org/10.1103/PhysRevD.50.2441}{\emph{Phys. Rev. D}
  {\bfseries 50} (1994) 2441}
  [\href{https://arxiv.org/abs/hep-ph/9311278}{{\ttfamily hep-ph/9311278}}].

\bibitem{Berera:1998gx}
A.~Berera, M.~Gleiser and R.O.~Ramos, \emph{{Strong dissipative behavior in
  quantum field theory}},
  \href{https://doi.org/10.1103/PhysRevD.58.123508}{\emph{Phys. Rev. D}
  {\bfseries 58} (1998) 123508}
  [\href{https://arxiv.org/abs/hep-ph/9803394}{{\ttfamily hep-ph/9803394}}].

\bibitem{Berera:2001gs}
A.~Berera and R.O.~Ramos, \emph{{The Affinity for scalar fields to dissipate}},
  \href{https://doi.org/10.1103/PhysRevD.63.103509}{\emph{Phys. Rev. D}
  {\bfseries 63} (2001) 103509}
  [\href{https://arxiv.org/abs/hep-ph/0101049}{{\ttfamily hep-ph/0101049}}].

\bibitem{Felder:2000hq}
G.N.~Felder and I.~Tkachev, \emph{{LATTICEEASY: A Program for lattice
  simulations of scalar fields in an expanding universe}},
  \href{https://doi.org/10.1016/j.cpc.2008.02.009}{\emph{Comput. Phys. Commun.}
  {\bfseries 178} (2008) 929}
  [\href{https://arxiv.org/abs/hep-ph/0011159}{{\ttfamily hep-ph/0011159}}].

\bibitem{Huang:2011gf}
Z.~Huang, \emph{{The Art of Lattice and Gravity Waves from Preheating}},
  \href{https://doi.org/10.1103/PhysRevD.83.123509}{\emph{Phys. Rev. D}
  {\bfseries 83} (2011) 123509}
  [\href{https://arxiv.org/abs/1102.0227}{{\ttfamily 1102.0227}}].

\bibitem{Figueroa:2021yhd}
D.G.~Figueroa, A.~Florio, F.~Torrenti and W.~Valkenburg, \emph{{CosmoLattice: A
  modern code for lattice simulations of scalar and gauge field dynamics in an
  expanding universe}},
  \href{https://doi.org/10.1016/j.cpc.2022.108586}{\emph{Comput. Phys. Commun.}
  {\bfseries 283} (2023) 108586}
  [\href{https://arxiv.org/abs/2102.01031}{{\ttfamily 2102.01031}}].

\bibitem{Berges:2002cz}
J.~Berges and J.~Serreau, \emph{{Parametric resonance in quantum field
  theory}}, \href{https://doi.org/10.1103/PhysRevLett.91.111601}{\emph{Phys.
  Rev. Lett.} {\bfseries 91} (2003) 111601}
  [\href{https://arxiv.org/abs/hep-ph/0208070}{{\ttfamily hep-ph/0208070}}].

\bibitem{Berges:2013lsa}
J.~Berges, K.~Boguslavski, S.~Schlichting and R.~Venugopalan, \emph{{Basin of
  attraction for turbulent thermalization and the range of validity of
  classical-statistical simulations}},
  \href{https://doi.org/10.1007/JHEP05(2014)054}{\emph{JHEP} {\bfseries 05}
  (2014) 054} [\href{https://arxiv.org/abs/1312.5216}{{\ttfamily 1312.5216}}].

\bibitem{Tranberg:2023uzs}
A.~Tranberg and G.~Ungersb\"ack, \emph{{Quantum tachyonic preheating,
  revisited}}, \href{https://doi.org/10.1007/JHEP05(2024)128}{\emph{JHEP}
  {\bfseries 05} (2024) 128}
  [\href{https://arxiv.org/abs/2312.08167}{{\ttfamily 2312.08167}}].

\bibitem{Jedamzik:2010dq}
K.~Jedamzik, M.~Lemoine and J.~Martin, \emph{{Collapse of Small-Scale Density
  Perturbations during Preheating in Single Field Inflation}},
  \href{https://doi.org/10.1088/1475-7516/2010/09/034}{\emph{JCAP} {\bfseries
  09} (2010) 034} [\href{https://arxiv.org/abs/1002.3039}{{\ttfamily
  1002.3039}}].

\bibitem{Easther:2010mr}
R.~Easther, R.~Flauger and J.B.~Gilmore, \emph{{Delayed Reheating and the
  Breakdown of Coherent Oscillations}},
  \href{https://doi.org/10.1088/1475-7516/2011/04/027}{\emph{JCAP} {\bfseries
  04} (2011) 027} [\href{https://arxiv.org/abs/1003.3011}{{\ttfamily
  1003.3011}}].

\bibitem{Khlebnikov:1996mc}
S.Y.~Khlebnikov and I.I.~Tkachev, \emph{{Classical decay of inflaton}},
  \href{https://doi.org/10.1103/PhysRevLett.77.219}{\emph{Phys. Rev. Lett.}
  {\bfseries 77} (1996) 219}
  [\href{https://arxiv.org/abs/hep-ph/9603378}{{\ttfamily hep-ph/9603378}}].

\bibitem{Greene:1997fu}
P.B.~Greene, L.~Kofman, A.D.~Linde and A.A.~Starobinsky, \emph{{Structure of
  resonance in preheating after inflation}},
  \href{https://doi.org/10.1103/PhysRevD.56.6175}{\emph{Phys. Rev. D}
  {\bfseries 56} (1997) 6175}
  [\href{https://arxiv.org/abs/hep-ph/9705347}{{\ttfamily hep-ph/9705347}}].

\bibitem{Micha:2002ey}
R.~Micha and I.I.~Tkachev, \emph{{Relativistic turbulence: A Long way from
  preheating to equilibrium}},
  \href{https://doi.org/10.1103/PhysRevLett.90.121301}{\emph{Phys. Rev. Lett.}
  {\bfseries 90} (2003) 121301}
  [\href{https://arxiv.org/abs/hep-ph/0210202}{{\ttfamily hep-ph/0210202}}].

\bibitem{Micha:2004bv}
R.~Micha and I.I.~Tkachev, \emph{{Turbulent thermalization}},
  \href{https://doi.org/10.1103/PhysRevD.70.043538}{\emph{Phys. Rev. D}
  {\bfseries 70} (2004) 043538}
  [\href{https://arxiv.org/abs/hep-ph/0403101}{{\ttfamily hep-ph/0403101}}].

\bibitem{Garcia:2023eol}
M.A.G.~Garc\'ia and M.~Pierre, \emph{{Reheating after inflaton fragmentation}},
  \href{https://doi.org/10.1088/1475-7516/2023/11/004}{\emph{JCAP} {\bfseries
  11} (2023) 004} [\href{https://arxiv.org/abs/2306.08038}{{\ttfamily
  2306.08038}}].

\bibitem{Garcia:2023dyf}
M.A.G.~Garc\'ia, M.~Gross, Y.~Mambrini, K.A.~Olive, M.~Pierre and J.-H.~Yoon,
  \emph{{Effects of fragmentation on post-inflationary reheating}},
  \href{https://doi.org/10.1088/1475-7516/2023/12/028}{\emph{JCAP} {\bfseries
  12} (2023) 028} [\href{https://arxiv.org/abs/2308.16231}{{\ttfamily
  2308.16231}}].

\bibitem{magnus2004hills}
W.~Magnus and S.~Winkler, \emph{Hill's Equation}, Dover Books on Mathematics,
  Dover Publications (2004).

\bibitem{Amin:2010xe}
M.A.~Amin, \emph{{Inflaton fragmentation: Emergence of pseudo-stable inflaton
  lumps (oscillons) after inflation}},
  \href{https://arxiv.org/abs/1006.3075}{{\ttfamily 1006.3075}}.

\bibitem{Amin:2010dc}
M.A.~Amin, R.~Easther and H.~Finkel, \emph{{Inflaton Fragmentation and Oscillon
  Formation in Three Dimensions}},
  \href{https://doi.org/10.1088/1475-7516/2010/12/001}{\emph{JCAP} {\bfseries
  12} (2010) 001} [\href{https://arxiv.org/abs/1009.2505}{{\ttfamily
  1009.2505}}].

\bibitem{Amin:2011hj}
M.A.~Amin, R.~Easther, H.~Finkel, R.~Flauger and M.P.~Hertzberg,
  \emph{{Oscillons After Inflation}},
  \href{https://doi.org/10.1103/PhysRevLett.108.241302}{\emph{Phys. Rev. Lett.}
  {\bfseries 108} (2012) 241302}
  [\href{https://arxiv.org/abs/1106.3335}{{\ttfamily 1106.3335}}].

\bibitem{Lozanov:2017hjm}
K.D.~Lozanov and M.A.~Amin, \emph{{Self-resonance after inflation: oscillons,
  transients and radiation domination}},
  \href{https://doi.org/10.1103/PhysRevD.97.023533}{\emph{Phys. Rev. D}
  {\bfseries 97} (2018) 023533}
  [\href{https://arxiv.org/abs/1710.06851}{{\ttfamily 1710.06851}}].

\bibitem{Piani:2023aof}
M.~Piani and J.~Rubio, \emph{{Preheating in Einstein-Cartan Higgs Inflation:
  oscillon formation}},
  \href{https://doi.org/10.1088/1475-7516/2023/12/002}{\emph{JCAP} {\bfseries
  12} (2023) 002} [\href{https://arxiv.org/abs/2304.13056}{{\ttfamily
  2304.13056}}].

\bibitem{Piani:2025dpy}
M.~Piani, J.~Rubio and F.~Torrenti, \emph{{Ephemeral Oscillons in Scalar-Tensor
  Theories: The Higgs-like case}},
  \href{https://arxiv.org/abs/2501.14869}{{\ttfamily 2501.14869}}.

\bibitem{Rubio:2019ypq}
J.~Rubio and E.S.~Tomberg, \emph{{Preheating in Palatini Higgs inflation}},
  \href{https://doi.org/10.1088/1475-7516/2019/04/021}{\emph{JCAP} {\bfseries
  04} (2019) 021} [\href{https://arxiv.org/abs/1902.10148}{{\ttfamily
  1902.10148}}].

\bibitem{Lozanov:2016hid}
K.D.~Lozanov and M.A.~Amin, \emph{{Equation of State and Duration to Radiation
  Domination after Inflation}},
  \href{https://doi.org/10.1103/PhysRevLett.119.061301}{\emph{Phys. Rev. Lett.}
  {\bfseries 119} (2017) 061301}
  [\href{https://arxiv.org/abs/1608.01213}{{\ttfamily 1608.01213}}].

\bibitem{Bettoni:2021zhq}
D.~Bettoni, A.~L\'opez-Eiguren and J.~Rubio, \emph{{Hubble-induced phase
  transitions on the lattice with applications to Ricci reheating}},
  \href{https://doi.org/10.1088/1475-7516/2022/01/002}{\emph{JCAP} {\bfseries
  01} (2022) 002} [\href{https://arxiv.org/abs/2107.09671}{{\ttfamily
  2107.09671}}].

\bibitem{Felder:2000hj}
G.N.~Felder, J.~García-Bellido, P.B.~Greene, L.~Kofman, A.D.~Linde and
  I.~Tkachev, \emph{{Dynamics of symmetry breaking and tachyonic preheating}},
  \href{https://doi.org/10.1103/PhysRevLett.87.011601}{\emph{Phys. Rev. Lett.}
  {\bfseries 87} (2001) 011601}
  [\href{https://arxiv.org/abs/hep-ph/0012142}{{\ttfamily hep-ph/0012142}}].

\bibitem{Felder:2001kt}
G.N.~Felder, L.~Kofman and A.D.~Linde, \emph{{Tachyonic instability and
  dynamics of spontaneous symmetry breaking}},
  \href{https://doi.org/10.1103/PhysRevD.64.123517}{\emph{Phys. Rev. D}
  {\bfseries 64} (2001) 123517}
  [\href{https://arxiv.org/abs/hep-th/0106179}{{\ttfamily hep-th/0106179}}].

\bibitem{Dufaux:2006ee}
J.F.~Dufaux, G.N.~Felder, L.~Kofman, M.~Peloso and D.~Podolsky,
  \emph{{Preheating with trilinear interactions: Tachyonic resonance}},
  \href{https://doi.org/10.1088/1475-7516/2006/07/006}{\emph{JCAP} {\bfseries
  07} (2006) 006} [\href{https://arxiv.org/abs/hep-ph/0602144}{{\ttfamily
  hep-ph/0602144}}].

\bibitem{Greene:1997ge}
B.R.~Greene, T.~Prokopec and T.G.~Roos, \emph{{Inflaton decay and heavy
  particle production with negative coupling}},
  \href{https://doi.org/10.1103/PhysRevD.56.6484}{\emph{Phys. Rev. D}
  {\bfseries 56} (1997) 6484}
  [\href{https://arxiv.org/abs/hep-ph/9705357}{{\ttfamily hep-ph/9705357}}].

\bibitem{Abolhasani:2009nb}
A.A.~Abolhasani, H.~Firouzjahi and M.M.~Sheikh-Jabbari, \emph{{Tachyonic
  Resonance Preheating in Expanding Universe}},
  \href{https://doi.org/10.1103/PhysRevD.81.043524}{\emph{Phys. Rev. D}
  {\bfseries 81} (2010) 043524}
  [\href{https://arxiv.org/abs/0912.1021}{{\ttfamily 0912.1021}}].

\bibitem{Khlebnikov:1996wr}
S.Y.~Khlebnikov and I.I.~Tkachev, \emph{{The Universe after inflation: The Wide
  resonance case}},
  \href{https://doi.org/10.1016/S0370-2693(96)01419-0}{\emph{Phys. Lett. B}
  {\bfseries 390} (1997) 80}
  [\href{https://arxiv.org/abs/hep-ph/9608458}{{\ttfamily hep-ph/9608458}}].

\bibitem{Khlebnikov:1996zt}
S.Y.~Khlebnikov and I.I.~Tkachev, \emph{{Resonant decay of Bose condensates}},
  \href{https://doi.org/10.1103/PhysRevLett.79.1607}{\emph{Phys. Rev. Lett.}
  {\bfseries 79} (1997) 1607}
  [\href{https://arxiv.org/abs/hep-ph/9610477}{{\ttfamily hep-ph/9610477}}].

\bibitem{Prokopec:1996rr}
T.~Prokopec and T.G.~Roos, \emph{{Lattice study of classical inflaton decay}},
  \href{https://doi.org/10.1103/PhysRevD.55.3768}{\emph{Phys. Rev. D}
  {\bfseries 55} (1997) 3768}
  [\href{https://arxiv.org/abs/hep-ph/9610400}{{\ttfamily hep-ph/9610400}}].

\bibitem{Battefeld:2008bu}
D.~Battefeld and S.~Kawai, \emph{{Preheating after N-flation}},
  \href{https://doi.org/10.1103/PhysRevD.77.123507}{\emph{Phys. Rev. D}
  {\bfseries 77} (2008) 123507}
  [\href{https://arxiv.org/abs/0803.0321}{{\ttfamily 0803.0321}}].

\bibitem{Battefeld:2009xw}
D.~Battefeld, T.~Battefeld and J.T.~Giblin, \emph{{On the Suppression of
  Parametric Resonance and the Viability of Tachyonic Preheating after
  Multi-Field Inflation}},
  \href{https://doi.org/10.1103/PhysRevD.79.123510}{\emph{Phys. Rev. D}
  {\bfseries 79} (2009) 123510}
  [\href{https://arxiv.org/abs/0904.2778}{{\ttfamily 0904.2778}}].

\bibitem{Greene:1998nh}
P.B.~Greene and L.~Kofman, \emph{{Preheating of fermions}},
  \href{https://doi.org/10.1016/S0370-2693(99)00020-9}{\emph{Phys. Lett. B}
  {\bfseries 448} (1999) 6}
  [\href{https://arxiv.org/abs/hep-ph/9807339}{{\ttfamily hep-ph/9807339}}].

\bibitem{Baacke:1998di}
J.~Baacke, K.~Heitmann and C.~Patzold, \emph{{Nonequilibrium dynamics of
  fermions in a spatially homogeneous scalar background field}},
  \href{https://doi.org/10.1103/PhysRevD.58.125013}{\emph{Phys. Rev. D}
  {\bfseries 58} (1998) 125013}
  [\href{https://arxiv.org/abs/hep-ph/9806205}{{\ttfamily hep-ph/9806205}}].

\bibitem{Greene:2000ew}
P.B.~Greene and L.~Kofman, \emph{{On the theory of fermionic preheating}},
  \href{https://doi.org/10.1103/PhysRevD.62.123516}{\emph{Phys. Rev. D}
  {\bfseries 62} (2000) 123516}
  [\href{https://arxiv.org/abs/hep-ph/0003018}{{\ttfamily hep-ph/0003018}}].

\bibitem{Peloso:2000hy}
M.~Peloso and L.~Sorbo, \emph{{Preheating of massive fermions after inflation:
  Analytical results}},
  \href{https://doi.org/10.1088/1126-6708/2000/05/016}{\emph{JHEP} {\bfseries
  05} (2000) 016} [\href{https://arxiv.org/abs/hep-ph/0003045}{{\ttfamily
  hep-ph/0003045}}].

\bibitem{Garcia-Bellido:2008ycs}
J.~Garc\'ia-Bellido, D.G.~Figueroa and J.~Rubio, \emph{{Preheating in the
  Standard Model with the Higgs-Inflaton coupled to gravity}},
  \href{https://doi.org/10.1103/PhysRevD.79.063531}{\emph{Phys. Rev. D}
  {\bfseries 79} (2009) 063531}
  [\href{https://arxiv.org/abs/0812.4624}{{\ttfamily 0812.4624}}].

\bibitem{Rubio:2015zia}
J.~Rubio, \emph{{Higgs inflation and vacuum stability}},
  \href{https://doi.org/10.1088/1742-6596/631/1/012032}{\emph{J. Phys. Conf.
  Ser.} {\bfseries 631} (2015) 012032}
  [\href{https://arxiv.org/abs/1502.07952}{{\ttfamily 1502.07952}}].

\bibitem{Repond:2016sol}
J.~Repond and J.~Rubio, \emph{{Combined Preheating on the lattice with
  applications to Higgs inflation}},
  \href{https://doi.org/10.1088/1475-7516/2016/07/043}{\emph{JCAP} {\bfseries
  07} (2016) 043} [\href{https://arxiv.org/abs/1604.08238}{{\ttfamily
  1604.08238}}].

\bibitem{Fan:2021otj}
J.~Fan, K.D.~Lozanov and Q.~Lu, \emph{{Spillway Preheating}},
  \href{https://doi.org/10.1007/JHEP05(2021)069}{\emph{JHEP} {\bfseries 05}
  (2021) 069} [\href{https://arxiv.org/abs/2101.11008}{{\ttfamily
  2101.11008}}].

\bibitem{Mansfield:2023sqp}
G.~Mansfield, J.~Fan and Q.~Lu, \emph{{Phenomenology of spillway preheating:
  Equation of state and gravitational waves}},
  \href{https://doi.org/10.1103/PhysRevD.110.023542}{\emph{Phys. Rev. D}
  {\bfseries 110} (2024) 023542}
  [\href{https://arxiv.org/abs/2312.03072}{{\ttfamily 2312.03072}}].

\bibitem{Bettoni:2019dcw}
D.~Bettoni and J.~Rubio, \emph{{Hubble-induced phase transitions: Walls are not
  forever}}, \href{https://doi.org/10.1088/1475-7516/2020/01/002}{\emph{JCAP}
  {\bfseries 01} (2020) 002}
  [\href{https://arxiv.org/abs/1911.03484}{{\ttfamily 1911.03484}}].

\bibitem{Laverda:2023uqv}
G.~Laverda and J.~Rubio, \emph{{Ricci reheating reloaded}},
  \href{https://doi.org/10.1088/1475-7516/2024/03/033}{\emph{JCAP} {\bfseries
  03} (2024) 033} [\href{https://arxiv.org/abs/2307.03774}{{\ttfamily
  2307.03774}}].

\bibitem{Bettoni:2018pbl}
D.~Bettoni, G.~Dom\`enech and J.~Rubio, \emph{{Gravitational waves from global
  cosmic strings in quintessential inflation}},
  \href{https://doi.org/10.1088/1475-7516/2019/02/034}{\emph{JCAP} {\bfseries
  02} (2019) 034} [\href{https://arxiv.org/abs/1810.11117}{{\ttfamily
  1810.11117}}].

\bibitem{Bettoni:2024ixe}
D.~Bettoni, G.~Laverda, A.L.~Eiguren and J.~Rubio, \emph{{Hubble-induced phase
  transitions: gravitational-wave imprint of Ricci reheating from lattice
  simulations}},
  \href{https://doi.org/10.1088/1475-7516/2025/03/027}{\emph{JCAP} {\bfseries
  03} (2025) 027} [\href{https://arxiv.org/abs/2409.15450}{{\ttfamily
  2409.15450}}].

\bibitem{Bettoni:2018utf}
D.~Bettoni and J.~Rubio, \emph{{Quintessential Affleck-Dine baryogenesis with
  non-minimal couplings}},
  \href{https://doi.org/10.1016/j.physletb.2018.07.046}{\emph{Phys. Lett. B}
  {\bfseries 784} (2018) 122}
  [\href{https://arxiv.org/abs/1805.02669}{{\ttfamily 1805.02669}}].

\bibitem{Chen:2025awt}
C.~Chen, S.~Jyoti~Das, K.~Dimopoulos and A.~Ghoshal, \emph{{Flipped Rotating
  Axion Non-minimally Coupled to Gravity: Baryogenesis and Dark Matter}},
  \href{https://arxiv.org/abs/2502.08720}{{\ttfamily 2502.08720}}.

\bibitem{Laverda:2024qjt}
G.~Laverda and J.~Rubio, \emph{{The rise and fall of the Standard-Model Higgs:
  electroweak vacuum stability during kination}},
  \href{https://doi.org/10.1007/JHEP05(2024)339}{\emph{JHEP} {\bfseries 05}
  (2024) 339} [\href{https://arxiv.org/abs/2402.06000}{{\ttfamily
  2402.06000}}].

\bibitem{Laverda:2025pmg}
G.~Laverda and J.~Rubio, \emph{{Higgs-Induced Gravitational Waves: the
  Interplay of Non-Minimal Couplings, Kination and Top Quark Mass}},
  \href{https://arxiv.org/abs/2502.04445}{{\ttfamily 2502.04445}}.

\bibitem{Shaposhnikov:1987tw}
M.E.~Shaposhnikov, \emph{{Baryon Asymmetry of the Universe in Standard
  Electroweak Theory}},
  \href{https://doi.org/10.1016/0550-3213(87)90127-1}{\emph{Nucl. Phys. B}
  {\bfseries 287} (1987) 757}.

\bibitem{Wagner:2023vqw}
C.E.M.~Wagner, \emph{{Electroweak Baryogenesis and Higgs Physics}},
  \href{https://doi.org/10.31526/lhep.2023.466}{\emph{LHEP} {\bfseries 2023}
  (2023) 466} [\href{https://arxiv.org/abs/2311.06949}{{\ttfamily
  2311.06949}}].

\bibitem{CMS:2023ebf}
{\scshape CMS} collaboration, \emph{{Measurement of the top quark mass using a
  profile likelihood approach with the lepton~+~jets final states in
  proton\textendash{}proton collisions at $\sqrt{s}=13\,\text
  {Te}\hspace{-.08em}\text {V} $}},
  \href{https://doi.org/10.1140/epjc/s10052-023-12050-4}{\emph{Eur. Phys. J. C}
  {\bfseries 83} (2023) 963}
  [\href{https://arxiv.org/abs/2302.01967}{{\ttfamily 2302.01967}}].

\bibitem{Myllymaki:2024uje}
{\scshape ATLAS, CMS} collaboration, \emph{{Top mass measurements}},  in
  \emph{{16$^{th}$ International Workshop on Top Quark Physics}}, 1, 2024
  [\href{https://arxiv.org/abs/2401.04824}{{\ttfamily 2401.04824}}].

\bibitem{Felder:2000hr}
G.N.~Felder and L.~Kofman, \emph{{The Development of equilibrium after
  preheating}}, \href{https://doi.org/10.1103/PhysRevD.63.103503}{\emph{Phys.
  Rev. D} {\bfseries 63} (2001) 103503}
  [\href{https://arxiv.org/abs/hep-ph/0011160}{{\ttfamily hep-ph/0011160}}].

\bibitem{Berges:2004ce}
J.~Berges, S.~Borsanyi and C.~Wetterich, \emph{{Prethermalization}},
  \href{https://doi.org/10.1103/PhysRevLett.93.142002}{\emph{Phys. Rev. Lett.}
  {\bfseries 93} (2004) 142002}
  [\href{https://arxiv.org/abs/hep-ph/0403234}{{\ttfamily hep-ph/0403234}}].

\bibitem{Podolsky:2005bw}
D.I.~Podolsky, G.N.~Felder, L.~Kofman and M.~Peloso, \emph{{Equation of state
  and beginning of thermalization after preheating}},
  \href{https://doi.org/10.1103/PhysRevD.73.023501}{\emph{Phys. Rev. D}
  {\bfseries 73} (2006) 023501}
  [\href{https://arxiv.org/abs/hep-ph/0507096}{{\ttfamily hep-ph/0507096}}].

\bibitem{Bernal:2020qyu}
N.~Bernal, J.~Rubio and H.~Veerm\"ae, \emph{{UV Freeze-in in Starobinsky
  Inflation}}, \href{https://doi.org/10.1088/1475-7516/2020/10/021}{\emph{JCAP}
  {\bfseries 10} (2020) 021}
  [\href{https://arxiv.org/abs/2006.02442}{{\ttfamily 2006.02442}}].

\bibitem{Stelle:1977ry}
K.S.~Stelle, \emph{{Classical Gravity with Higher Derivatives}},
  \href{https://doi.org/10.1007/BF00760427}{\emph{Gen. Rel. Grav.} {\bfseries
  9} (1978) 353}.

\bibitem{Whitt:1984pd}
B.~Whitt, \emph{{Fourth Order Gravity as General Relativity Plus Matter}},
  \href{https://doi.org/10.1016/0370-2693(84)90332-0}{\emph{Phys. Lett. B}
  {\bfseries 145} (1984) 176}.

\bibitem{Mukhanov:1989rq}
V.F.~Mukhanov, \emph{{Quantum Theory of Cosmological Perturbations in $R^2$
  Gravity}}, \href{https://doi.org/10.1016/0370-2693(89)90467-X}{\emph{Phys.
  Lett. B} {\bfseries 218} (1989) 17}.

\bibitem{Takeda:2014qma}
N.~Takeda and Y.~Watanabe, \emph{{No quasistable scalaron lump forms after
  $R^2$ inflation}},
  \href{https://doi.org/10.1103/PhysRevD.90.023519}{\emph{Phys. Rev. D}
  {\bfseries 90} (2014) 023519}
  [\href{https://arxiv.org/abs/1405.3830}{{\ttfamily 1405.3830}}].

\bibitem{Watanabe:2010vy}
Y.~Watanabe, \emph{{Rate of gravitational inflaton decay via gauge trace
  anomaly}}, \href{https://doi.org/10.1103/PhysRevD.83.043511}{\emph{Phys. Rev.
  D} {\bfseries 83} (2011) 043511}
  [\href{https://arxiv.org/abs/1011.3348}{{\ttfamily 1011.3348}}].

\bibitem{Watanabe:2006ku}
Y.~Watanabe and E.~Komatsu, \emph{{Reheating of the universe after inflation
  with $f(\phi) R$ gravity}},
  \href{https://doi.org/10.1103/PhysRevD.75.061301}{\emph{Phys. Rev. D}
  {\bfseries 75} (2007) 061301}
  [\href{https://arxiv.org/abs/gr-qc/0612120}{{\ttfamily gr-qc/0612120}}].

\bibitem{Rudenok:2014daa}
I.~Rudenok, Y.~Shtanov and S.~Vilchinskii, \emph{{Post-inflationary preheating
  with weak coupling}},
  \href{https://doi.org/10.1016/j.physletb.2014.04.046}{\emph{Phys. Lett. B}
  {\bfseries 733} (2014) 193}
  [\href{https://arxiv.org/abs/1401.7298}{{\ttfamily 1401.7298}}].

\bibitem{Ellis:2015jpg}
J.~Ellis, M.A.G.~Garc\'ia, D.V.~Nanopoulos, K.A.~Olive and M.~Peloso,
  \emph{{Post-Inflationary Gravitino Production Revisited}},
  \href{https://doi.org/10.1088/1475-7516/2016/03/008}{\emph{JCAP} {\bfseries
  03} (2016) 008} [\href{https://arxiv.org/abs/1512.05701}{{\ttfamily
  1512.05701}}].

\bibitem{Garcia:2018wtq}
M.A.G.~Garc\'ia and M.A.~Amin, \emph{{Prethermalization production of dark
  matter}}, \href{https://doi.org/10.1103/PhysRevD.98.103504}{\emph{Phys. Rev.
  D} {\bfseries 98} (2018) 103504}
  [\href{https://arxiv.org/abs/1806.01865}{{\ttfamily 1806.01865}}].

\bibitem{Gorbunov:2010bn}
D.S.~Gorbunov and A.G.~Panin, \emph{{Scalaron the mighty: producing dark matter
  and baryon asymmetry at reheating}},
  \href{https://doi.org/10.1016/j.physletb.2011.04.067}{\emph{Phys. Lett. B}
  {\bfseries 700} (2011) 157}
  [\href{https://arxiv.org/abs/1009.2448}{{\ttfamily 1009.2448}}].

\bibitem{Gorbunov:2012ns}
D.~Gorbunov and A.~Tokareva, \emph{{$R^2$-inflation with conformal SM Higgs
  field}}, \href{https://doi.org/10.1088/1475-7516/2013/12/021}{\emph{JCAP}
  {\bfseries 12} (2013) 021} [\href{https://arxiv.org/abs/1212.4466}{{\ttfamily
  1212.4466}}].

\bibitem{Felder:1998vq}
G.N.~Felder, L.~Kofman and A.D.~Linde, \emph{{Instant preheating}},
  \href{https://doi.org/10.1103/PhysRevD.59.123523}{\emph{Phys. Rev. D}
  {\bfseries 59} (1999) 123523}
  [\href{https://arxiv.org/abs/hep-ph/9812289}{{\ttfamily hep-ph/9812289}}].

\bibitem{Felder:1999pv}
G.N.~Felder, L.~Kofman and A.D.~Linde, \emph{{Inflation and preheating in NO
  models}}, \href{https://doi.org/10.1103/PhysRevD.60.103505}{\emph{Phys. Rev.
  D} {\bfseries 60} (1999) 103505}
  [\href{https://arxiv.org/abs/hep-ph/9903350}{{\ttfamily hep-ph/9903350}}].

\bibitem{Rubio:2017gty}
J.~Rubio and C.~Wetterich, \emph{{Emergent scale symmetry: Connecting inflation
  and dark energy}},
  \href{https://doi.org/10.1103/PhysRevD.96.063509}{\emph{Phys. Rev. D}
  {\bfseries 96} (2017) 063509}
  [\href{https://arxiv.org/abs/1705.00552}{{\ttfamily 1705.00552}}].

\bibitem{Wetterich:2014gaa}
C.~Wetterich, \emph{{Inflation, quintessence, and the origin of mass}},
  \href{https://doi.org/10.1016/j.nuclphysb.2015.05.019}{\emph{Nucl. Phys. B}
  {\bfseries 897} (2015) 111}
  [\href{https://arxiv.org/abs/1408.0156}{{\ttfamily 1408.0156}}].

\bibitem{Bettoni:2021qfs}
D.~Bettoni and J.~Rubio, \emph{{Quintessential Inflation: A Tale of Emergent
  and Broken Symmetries}},
  \href{https://doi.org/10.3390/galaxies10010022}{\emph{Galaxies} {\bfseries
  10} (2022) 22} [\href{https://arxiv.org/abs/2112.11948}{{\ttfamily
  2112.11948}}].

\bibitem{Wetterich:2013jsa}
C.~Wetterich, \emph{{Variable gravity Universe}},
  \href{https://doi.org/10.1103/PhysRevD.89.024005}{\emph{Phys. Rev. D}
  {\bfseries 89} (2014) 024005}
  [\href{https://arxiv.org/abs/1308.1019}{{\ttfamily 1308.1019}}].

\bibitem{Uzan:2010pm}
J.-P.~Uzan, \emph{{Varying Constants, Gravitation and Cosmology}},
  \href{https://doi.org/10.12942/lrr-2011-2}{\emph{Living Rev. Rel.} {\bfseries
  14} (2011) 2} [\href{https://arxiv.org/abs/1009.5514}{{\ttfamily
  1009.5514}}].

\bibitem{Wetterich:2003qb}
C.~Wetterich, \emph{{Cosmology with varying scales and couplings}},  in
  \emph{{$5^{th}$ Internationa Conference on Strong and Electroweak Matter}},
  pp.~230--249, 2003, \href{https://doi.org/10.1142/9789812704498_0022}{DOI}
  [\href{https://arxiv.org/abs/hep-ph/0302116}{{\ttfamily hep-ph/0302116}}].

\bibitem{Bernal:2020bfj}
N.~Bernal, J.~Rubio and H.~Veerm\"ae, \emph{{Boosting Ultraviolet Freeze-in in
  NO Models}}, \href{https://doi.org/10.1088/1475-7516/2020/06/047}{\emph{JCAP}
  {\bfseries 06} (2020) 047}
  [\href{https://arxiv.org/abs/2004.13706}{{\ttfamily 2004.13706}}].

\bibitem{Caprini:2018mtu}
C.~Caprini and D.G.~Figueroa, \emph{{Cosmological Backgrounds of Gravitational
  Waves}}, \href{https://doi.org/10.1088/1361-6382/aac608}{\emph{Class. Quant.
  Grav.} {\bfseries 35} (2018) 163001}
  [\href{https://arxiv.org/abs/1801.04268}{{\ttfamily 1801.04268}}].

\bibitem{Kolb:1998he}
E.W.~Kolb, A.~Riotto and I.I.~Tkachev, \emph{{GUT baryogenesis after
  preheating: Numerical study of the production and decay of X bosons}},
  \href{https://doi.org/10.1016/S0370-2693(98)00134-8}{\emph{Phys. Lett. B}
  {\bfseries 423} (1998) 348}
  [\href{https://arxiv.org/abs/hep-ph/9801306}{{\ttfamily hep-ph/9801306}}].

\bibitem{Giudice:1999fb}
G.F.~Giudice, M.~Peloso, A.~Riotto and I.~Tkachev, \emph{{Production of massive
  fermions at preheating and leptogenesis}},
  \href{https://doi.org/10.1088/1126-6708/1999/08/014}{\emph{JHEP} {\bfseries
  08} (1999) 014} [\href{https://arxiv.org/abs/hep-ph/9905242}{{\ttfamily
  hep-ph/9905242}}].

\bibitem{Krauss:1999ng}
L.M.~Krauss and M.~Trodden, \emph{{Baryogenesis below the electroweak scale}},
  \href{https://doi.org/10.1103/PhysRevLett.83.1502}{\emph{Phys. Rev. Lett.}
  {\bfseries 83} (1999) 1502}
  [\href{https://arxiv.org/abs/hep-ph/9902420}{{\ttfamily hep-ph/9902420}}].

\bibitem{Garcia-Bellido:1999xos}
J.~García-Bellido, D.Y.~Grigoriev, A.~Kusenko and M.E.~Shaposhnikov,
  \emph{{Nonequilibrium electroweak baryogenesis from preheating after
  inflation}}, \href{https://doi.org/10.1103/PhysRevD.60.123504}{\emph{Phys.
  Rev. D} {\bfseries 60} (1999) 123504}
  [\href{https://arxiv.org/abs/hep-ph/9902449}{{\ttfamily hep-ph/9902449}}].

\bibitem{Copeland:2001qw}
E.J.~Copeland, D.~Lyth, A.~Rajantie and M.~Trodden, \emph{{Hybrid inflation and
  baryogenesis at the TeV scale}},
  \href{https://doi.org/10.1103/PhysRevD.64.043506}{\emph{Phys. Rev. D}
  {\bfseries 64} (2001) 043506}
  [\href{https://arxiv.org/abs/hep-ph/0103231}{{\ttfamily hep-ph/0103231}}].

\bibitem{Garcia-Bellido:2003wva}
J.~García-Bellido, M.~García-Perez and A.~González-Arroyo,
  \emph{{Chern-Simons production during preheating in hybrid inflation
  models}}, \href{https://doi.org/10.1103/PhysRevD.69.023504}{\emph{Phys. Rev.
  D} {\bfseries 69} (2004) 023504}
  [\href{https://arxiv.org/abs/hep-ph/0304285}{{\ttfamily hep-ph/0304285}}].

\bibitem{Tranberg:2006dg}
A.~Tranberg, J.~Smit and M.~Hindmarsh, \emph{{Simulations of cold electroweak
  baryogenesis: Finite time quenches}},
  \href{https://doi.org/10.1088/1126-6708/2007/01/034}{\emph{JHEP} {\bfseries
  01} (2007) 034} [\href{https://arxiv.org/abs/hep-ph/0610096}{{\ttfamily
  hep-ph/0610096}}].

\bibitem{Bassett:1998wg}
B.A.~Bassett, D.I.~Kaiser and R.~Maartens, \emph{{General relativistic
  preheating after inflation}},
  \href{https://doi.org/10.1016/S0370-2693(99)00478-5}{\emph{Phys. Lett. B}
  {\bfseries 455} (1999) 84}
  [\href{https://arxiv.org/abs/hep-ph/9808404}{{\ttfamily hep-ph/9808404}}].

\bibitem{Finelli:1998bu}
F.~Finelli and R.H.~Brandenberger, \emph{{Parametric amplification of
  gravitational fluctuations during reheating}},
  \href{https://doi.org/10.1103/PhysRevLett.82.1362}{\emph{Phys. Rev. Lett.}
  {\bfseries 82} (1999) 1362}
  [\href{https://arxiv.org/abs/hep-ph/9809490}{{\ttfamily hep-ph/9809490}}].

\bibitem{Chambers:2007se}
A.~Chambers and A.~Rajantie, \emph{{Lattice calculation of non-Gaussianity from
  preheating}},
  \href{https://doi.org/10.1103/PhysRevLett.100.041302}{\emph{Phys. Rev. Lett.}
  {\bfseries 100} (2008) 041302}
  [\href{https://arxiv.org/abs/0710.4133}{{\ttfamily 0710.4133}}].

\bibitem{Bond:2009xx}
J.R.~Bond, A.V.~Frolov, Z.~Huang and L.~Kofman, \emph{{Non-Gaussian Spikes from
  Chaotic Billiards in Inflation Preheating}},
  \href{https://doi.org/10.1103/PhysRevLett.103.071301}{\emph{Phys. Rev. Lett.}
  {\bfseries 103} (2009) 071301}
  [\href{https://arxiv.org/abs/0903.3407}{{\ttfamily 0903.3407}}].

\bibitem{Imrith:2019njf}
S.V.~Imrith, D.J.~Mulryne and A.~Rajantie, \emph{{Primordial curvature
  perturbation from lattice simulations}},
  \href{https://doi.org/10.1103/PhysRevD.100.043543}{\emph{Phys. Rev. D}
  {\bfseries 100} (2019) 043543}
  [\href{https://arxiv.org/abs/1903.07487}{{\ttfamily 1903.07487}}].

\bibitem{Giblin:2019nuv}
J.T.~Giblin and A.J.~Tishue, \emph{{Preheating in Full General Relativity}},
  \href{https://doi.org/10.1103/PhysRevD.100.063543}{\emph{Phys. Rev. D}
  {\bfseries 100} (2019) 063543}
  [\href{https://arxiv.org/abs/1907.10601}{{\ttfamily 1907.10601}}].

\bibitem{Adshead:2023mvt}
P.~Adshead, J.T.~Giblin, R.~Grutkoski and Z.J.~Weiner, \emph{{Gauge preheating
  with full general relativity}},
  \href{https://doi.org/10.1088/1475-7516/2024/03/017}{\emph{JCAP} {\bfseries
  03} (2024) 017} [\href{https://arxiv.org/abs/2311.01504}{{\ttfamily
  2311.01504}}].

\bibitem{Kasuya:1998aq}
S.~Kasuya and M.~Kawasaki, \emph{{Formation of topological defects during
  preheating}}, \href{https://doi.org/10.1063/1.59444}{\emph{AIP Conf. Proc.}
  {\bfseries 478} (1999) 75}.

\bibitem{Rajantie:2000fd}
A.~Rajantie and E.J.~Copeland, \emph{{Phase transitions from preheating in
  gauge theories}},
  \href{https://doi.org/10.1103/PhysRevLett.85.916}{\emph{Phys. Rev. Lett.}
  {\bfseries 85} (2000) 916}
  [\href{https://arxiv.org/abs/hep-ph/0003025}{{\ttfamily hep-ph/0003025}}].

\bibitem{Copeland:2002ku}
E.J.~Copeland, S.~Pascoli and A.~Rajantie, \emph{{Dynamics of tachyonic
  preheating after hybrid inflation}},
  \href{https://doi.org/10.1103/PhysRevD.65.103517}{\emph{Phys. Rev. D}
  {\bfseries 65} (2002) 103517}
  [\href{https://arxiv.org/abs/hep-ph/0202031}{{\ttfamily hep-ph/0202031}}].

\bibitem{Dufaux:2010cf}
J.-F.~Dufaux, D.G.~Figueroa and J.~García-Bellido, \emph{{Gravitational Waves
  from Abelian Gauge Fields and Cosmic Strings at Preheating}},
  \href{https://doi.org/10.1103/PhysRevD.82.083518}{\emph{Phys. Rev. D}
  {\bfseries 82} (2010) 083518}
  [\href{https://arxiv.org/abs/1006.0217}{{\ttfamily 1006.0217}}].

\end{thebibliography}\endgroup
\end{document}